\documentclass[10pt]{article}
\usepackage{amsmath}
\usepackage{graphicx}
\usepackage{amsfonts}
\usepackage{amssymb}
\usepackage{epsf}
\usepackage{latexsym}

\def\80{\hspace{0.8in}}

\newcommand{\bi}{\begin{itemize}}
\newcommand{\ei}{\end{itemize}}
\newcommand{\bd}{\begin{description}}
\newcommand{\ed}{\end{description}}
\def\beq{\begin{equation}}
\def\eeq{\end{equation}}
\def\bea{\begin{eqnarray}}
\def\eea{\end{eqnarray}}
\def\hat{\widehat}

\def\pa{\partial}
\def\d{\textrm{d}}

\def\cr{\mbox{\scriptsize{\bf $\mbox{ } \times \mbox{ }$}}}

\def\ma{\mbox{a}}

\def\mh{\mbox{h}}

\def\mj{\mbox{j}} 

\def\ml{\mbox{l}}
\def\mm{\mbox{m}}
\def\mn{\mbox{n}}
\def\mo{\mbox{o}}

\def\mr{\mbox{r}}

\def\mt{\mbox{t}}

\def\mC{\mbox{C}}
\def\mD{\mbox{D}}
\def\mE{\mbox{E}}

\def\mI{\mbox{I}}
\def\mJ{\mbox{J}}

\def\mL{\mbox{L}}

\def\mN{\mbox{N}} 

\def\mP{\mbox{P}}

\def\mR{\mbox{R}}

\def\sa{\mbox{\scriptsize a}}

\def\sc{\mbox{\scriptsize c}}
\def\sd{\mbox{\scriptsize d}}
\def\se{\mbox{\scriptsize e}}
\def\sf{\mbox{\scriptsize f}}
 
\def\sh{\mbox{\scriptsize h}} 
\def\si{\mbox{\scriptsize i}}
\def\sj{\mbox{\scriptsize j}} 

\def\sll{\mbox{\scriptsize l}}  
\def\sm{\mbox{\scriptsize m}}
\def\sn{\mbox{\scriptsize n}} 
\def\so{\mbox{\scriptsize o}}

\def\sr{\mbox{\scriptsize r}}
\def\sss{\mbox{\scriptsize s}}
\def\st{\mbox{\scriptsize t}}
\def\su{\mbox{\scriptsize u}}

\def\sw{\mbox{\scriptsize w}}

\def\sD{\mbox{\scriptsize D}}
\def\sE{\mbox{\scriptsize E}}

\def\sN{\mbox{\scriptsize N}}

\def\sR{\mbox{\scriptsize R}}

\def\barp{\bar{p}}

\def\eph(B){\mbox{\scriptsize emergent(LMB)}}

\def\tc{\mbox{\tiny c}}

\def\te{\mbox{\tiny e}}

\def\tm{\mbox{\tiny m}}

\def\tr{\mbox{\tiny r}}

\def\fE{\mbox{\sffamily E}}

\def\fH{\mbox{\sffamily H}}

\def\fQ{\mbox{\sffamily Q}}
\def\fR{\mbox{\sffamily R}}
\def\fS{\mbox{\sffamily S}}
\def\fT{\mbox{\sffamily T}}

\renewcommand{\H}{{\cal H}}                 % !!! Command redefined !!!

\def\brho{\mbox{\boldmath$\rho$}}

\begin{document}

\begin{titlepage}
\vspace{.7in}
\begin{center}

\LARGE{\bf SCALED TRIANGLELAND MODEL}

\vspace{.2in}

\LARGE{\bf OF QUANTUM COSMOLOGY} 

\vspace{.4in}

\large{\bf Edward Anderson}$^{1}$

\vspace{.2in}

\large{\em DAMTP, Centre for Mathematical Sciences, Wilberforce Road, Cambridge CB3 OWA.}

\end{center}

\begin{abstract}

In scaled relational particle mechanics, only relative times, relative angles and relative separations 
are meaningful. 
It arose in the study of the absolute versus relative motion debate. 
It has then turned out to be a useful toy model of classical and quantum general 
relativity, such as for investigating conceptual strategies for the problem of time. 
This paper studies the 3-particle 2-$d$ scaled relational particle model, for which the configurations  
are scaled triangles. 
The configuration space for these is $\mathbb{R}^3$ with a conformally flat metric thereupon 
(it is the cone over the corresponding shape space $\mathbb{S}^2$).   
I use multiple harmonic oscillator type potentials and other potentials suggested by analogy with 
cosmology, and solve for some of these by using a partial analogy with the treatment of the atom in 
spherical and parabolic coordinates.  
Spherical coordinates are here the total moment of inertia $I$ for radius and two pure-shape 
coordinates.  
These are $\Theta$, a function of the ratio of the two relative separations of subsystems, and 
$\Phi$, the relative angle between the two subsystems.  
Parabolic coordinates are $\Phi$ again and twice the partial moments of inertia of each subsystem.   
I interpret these solutions using 1) a `Bohr moment of inertia' for the model universe (playing the 
role of the scalefactor). 
2) Expectations and spreads of sizes and shapes. 
3) Superimposing the probability density function on the labelled tessellation of the configuration 
space that encodes meaningful subregions such as collinear configurations, equilateral triangles and 
isosceles triangles.
Applications include hidden time, emergent semiclassical time, timeless and histories theory problem of 
time strategies, and comparing reduced and Dirac methods of quantization.  

\end{abstract}

\vspace{1in}


\vspace{1in}


\mbox{ }

\vspace{1in}

\noindent$^1$ ea212@cam.ac.uk

\end{titlepage}

%========================================================================================================
%========================================================================================================
\section{Introduction}
%========================================================================================================
%========================================================================================================

%========================================================================================================
\subsection{What are Relational Particle Mechanics models?}
%========================================================================================================

{\it Scaled relational particle mechanics (RPM)} is a mechanics in which only relative times, 
relative angles and relative separations are physically meaningful.  
It was originally proposed in \cite{BB82} and further studied in \cite{ERPM, B94I, EOT, 06I, TriCl, 
08I, Cones, ScaleQM}.
{\it Pure-shape RPM} is a mechanics in which only relative times, 
relative angles and ratios of relative separations are meaningful.
It was originally proposed in \cite{B03} and further studied in \cite{SRPM, 06II, TriCl, FORD, 08I, 
08II, +tri, AF, Forth}.  
Note that these theories are relational in Barbour's sense of the word (which is more specific than 
Rovelli's distinct sense of the word, c.f. \cite{Rovellibook, B94I, EOT, 08I}), which involves   
the following postulates and implementations.   

\noindent 1) These theories are {\it temporally relational} \cite{BB82, RWR, Lan, FORD}. 
I.e., there is no meaningful primary notion of time for the whole system (e.g. the universe). 
This is mathematically implemented by using actions that are manifestly reparametrization 
invariant while also being free of extraneous\footnote{Relational programs should perhaps also steer 
%%%%%%%%%%%%%%%%%%%%%%%%%%%%%%%%%%%%%%%%%%%%%%%%%%%%%%%%%%%%%%%%%%%%%%%%%%%%%%%%%%%%%%%%%%%%%%%%%%%%%%%%%
clear of extraneous spatial/configurational structures such as background metrics.} 
%%%%%%%%%%%%%%%%%%%%%%%%%%%%%%%%%%%%%%%%%%%%%%%%%%%%%%%%%%%%%%%%%%%%%%%%%%%%%%%%%%%%%%%%%%%%%%%%%%%%%%%%%
time-related variables [e.g. Newtonian time or General Relativity (GR)'s lapse].   
This reparametrization invariance then directly produces primary constraints quadratic in the momenta 
\cite{Dirac}. 

\noindent 2) These theories are {\it configurationally relational}. 
This can be thought of in terms of a certain group $G$ of transformations that act on the 
theory's configuration space $\fQ$ leaving the physical configuration unchanged \cite{BB82, RWR, Lan, FORD, 
FEPI, Cones}.   
For scaled RPM, $G$ is the Euclidean group of translations and rotations and for pure-shape RPM it is the 
similarity group of translations, rotations and dilations.  
This can be implemented by such as using arbitrary-$G$-frame-corrected quantities rather than `bare' 
$\fQ$-configurations.
While this augments $\fQ$ to the principal bundle $P(\fQ, G)$, subsequent variation with respect to each 
adjoined independent auxiliary $G$-variable produces a secondary constraint linear in the momenta. 
This then removes both one $G$ degree of freedom and one redundant degree of freedom from $\fQ$. 
Thus one does end up on the desired reduced configuration space -- the quotient space $\fQ/G$.  
Configurational relationalism includes both spatial relationalism (for spatial transformations) and 
internal relationalism (in the sense of gauge theory).

%========================================================================================================
\subsection{Motivation for RPM's: toy models for classical and quantum GR}
%========================================================================================================

My principal motivation\footnote{Studying
%%%%%%%%%%%%%%%%%%%%%%%%%%%%%%%%%%%%%%%%%%%%%%%%%%%%%%%%%%%%%%%%%%%%%%%%%%%%%%%%%%%%%%%%%%%%%%%%%%%%%%%%%
RPM's has previously been motivated also by the long-standing absolute versus relative (relational) 
motion debate \cite{AORM, +Phil}, or from RPM's making useful examples in the study of quantization 
techniques \cite{BS89etc, Forth}. }
%%%%%%%%%%%%%%%%%%%%%%%%%%%%%%%%%%%%%%%%%%%%%%%%%%%%%%%%%%%%%%%%%%%%%%%%%%%%%%%%%%%%%%%%%%%%%%%%%%%%%%%%%
\cite{K92, 06II, AF, Cones} for studying RPM's is that they are useful toy models of GR its formulation as  
`geometrodynamics' (evolving spatial geometries).  
This analogy is quite rich (\cite{Cones} has a more detailed account). 
The extent of the resemblance (particularly in the formulations \cite{BSW, RWR, ABFKO} of GR) is  
comparable but different to the resemblance between GR and the more habitually studied minisuperspace 
models \cite{Mini, Magic, HH83, Wiltshire}.
RPM's have a quadratic energy constraint $H$ that is analogous to GR's quadratic Hamiltonian 
constraint ${\cal H}$.   
RPM's have a linear zero total angular momentum constraint $L_{a}$ is a nontrivial analogue of 
GR's linear momentum constraint ${\cal L}_{a}$ (for $a$ a spatial index).  
This is a structure which minisuperspace only possesses in a trivial sense, and yet it is important 
for a large number of detailed applications, some of which are discussed below.   
Also RPM's (unlike minisuperspace) have notions of locality in space and thus of clustering/structure.   
This is e.g. useful for cosmological modelling.

RPM's have many further useful analogies \cite{K92, B94I, B94II, EOT, Paris, 06II, SemiclI, Records, 
08II, AF, Cones} as regards conceptual aspects of Quantum Cosmology, including the Problem of Time 
\cite{K81, K91, K92, I93, K99, Kieferbook, Smolin08, APOT}.
This notorious problem occurs because `time' takes a different meaning in each of GR and ordinary 
quantum theory.  
This incompatibility underscores a number of problems with trying to replace these two branches with a 
single framework in situations in which the premises of both apply, such as in black holes or the very 
early universe.  
One facet of the Problem of Time appears in attempting canonical quantization of GR due to ${\cal H}$ 
being quadratic but not linear in the momenta. 
Then elevating ${\cal H}$ to a quantum equation produces a stationary i.e timeless or frozen wave 
equation -- the  Wheeler-DeWitt equation 
\beq
\hat{\cal H}\Psi = 0
\eeq 
(for $\Psi$ the wavefunction of the Universe) -- instead of ordinary QM's time-dependent one, 
\beq
i\hbar\pa\Psi/\pa t = \hat{\H}\Psi
\eeq 
(where I use $H$ to denote a Hamiltonian and $t$ absolute Newtonian time).  
See \cite{K92, I93} for other facets of the Problem of Time.

Some of the strategies toward resolving the Problem of Time are as follows.  

\noindent A) Perhaps one is to find a hidden time at the classical level \cite{K92} by extending or 
rearranging (1) to $p_{t^{\mbox{\tiny hidden}}  } + \fH_{\st\sr\su\se} = 0$ (for 
$p_{t^{\mbox{\tiny hidden}}}$ the momentum conjugate to some new coordinate $t^{\sh\si\sd\sd\se\sn}$ 
that is a candidate hidden timefunction).  
One would then promote this to a hidden-time-dependent Schr\"{o}dinger equation 
\beq
i\hbar\pa\Psi/\pa t^{\mbox{\tiny hidden}} = \hat{H}_{\st\sr\su\se}\Psi \mbox{ } .
\eeq
York time \cite{York72, York73, K81, K92, I93} is a candidate hidden time for GR.  
It is associated 
with constant mean curvature foliations \cite{York73, BS03} and corresponding conformal superspace 
(pure shape) formulations of GR \cite{York73, York74, ABFO, ABFKO}.

\noindent B) Perhaps one has slow heavy `$h$'  variables that provide an approximate timestandard with 
respect to which the other fast light `$l$' degrees of freedom evolve \cite{HallHaw, K92, Kieferbook}.  
In the Halliwell--Hawking \cite{HallHaw} scheme for GR Quantum Cosmology, $h$ is scale (and homogeneous 
matter modes) and $l$ are small inhomogeneities.
Thus the scale--shape split of scaled RPM's afford a tighter parallel of this \cite{MGM, Forth} than 
pure-shape RPM's.    
The semiclassical approach involves firstly making the Born--Oppenheimer ansatz $\Psi(h, l) = \psi(\mh)
|\chi(h, l) \rangle$ and the WKB ansatz $\psi(h) = \mbox{exp}(iW(h)/\hbar)$.  
Secondly, one forms the $h$-equation ($\langle\chi| \hat{H} \Psi = 0$ for RPM's), which, under a number 
of simplifications, yields a Hamilton--Jacobi\footnote{For simplicity, this is
%%%%%%%%%%%%%%%%%%%%%%%%%%%%%%%%%%%%%%%%%%%%%%%%%%%%%%%%%%%%%%%%%%%%%%%%%%%%%%%%%%%%%%%%%%%%%%%%%%%%%%%%% 
presented in the case of 1 $h$ degree of freedom and with no linear constraints.} 
%%%%%%%%%%%%%%%%%%%%%%%%%%%%%%%%%%%%%%%%%%%%%%%%%%%%%%%%%%%%%%%%%%%%%%%%%%%%%%%%%%%%%%%%%%%%%%%%%%%%%%%%%
equation
\beq
(\pa W/\pa h)^2 = 2(E - V(h)) \mbox{ } 
\label{HamJac} 
\eeq
for $E$ the total energy and $V(h)$ the $h$-part of the potential. 
Thirdly, one way of solving this is for an approximate emergent semiclassical time 
$t^{\se\sm} = t^{\se\sm}(h)$. 
Next, the $l$-equation $(1 - |\chi\rangle\langle\chi|)\hat{\H}\Psi = 0$ can be recast (modulo further 
approximations) into an emergent-time-dependent Schr\"{o}dinger equation for the $l$ degrees of freedom
\beq
i\hbar\pa|\chi\rangle/\pa t^{\te\tm}  = \widehat{H}_{l}|\chi\rangle \mbox{ } .  
\label{TDSE2}
\eeq
(Here the left-hand side arises from the cross-term $\pa_{h}|\chi\rangle\pa_{h}\psi$ and 
$\widehat{H}_{l}$ is the remaining surviving piece of $\widehat{H}$).  

\noindent C) A number of approaches take timelessness at face value. 
One considers only questions about the universe `being', rather than `becoming', a certain way.  
This has at least some practical limitations, but can address some questions of interest. 
Example 1: the {\it na\"{\i}ve Schr\"{o}dinger interpretation} \cite{HP86,UW89} concerns the `being' 
probabilities for universe properties such as: what is the probability that the universe is large? 
Flat? 
Isotropic? 
Homogeneous?   
One obtains these via consideration of the probability that the universe belongs to region $R$ of the 
configuration space that corresponds to a quantification of a particular such property, 
$P(R) \propto \int_{R}|\Psi|^2\d\Omega$, for $\d\Omega$ the configuration space volume element.
This approach is termed `na\"{\i}ve' due to it not using any further features of the constraint 
equations.  
Example 2: the {\it conditional probabilities interpretation} \cite{PW83} goes further by addressing 
conditioned questions of `being' such as `what is the probability that the universe is flat given that 
it is isotropic'?  
Example 3: {\it records theory} \cite{PW83, GMH, B94II, EOT, H99, Records} involves localized subconfigurations 
of a single instant.  
More concretely, it concerns whether these contain useable information, are correlated to each other, 
and a semblance of dynamics or history arises from this.  
This requires notions of localization in space and in configuration space as well as notions of 
information.  
RPM's are superior to minisuperspace for such a study as, firstly, they have a notion of localization 
in space. 
Secondly, they have more options for well-characterized localization in configuration space (i.e. of 
`distance between two shapes' \cite{Forth}) through their kinetic terms possessing positive-definite 
metrics.  

\noindent D) Perhaps instead it is the histories that are primary ({\it histories theory} \cite{GMH, 
Hartle}).    

\noindent E) Another current program worth mentioning is a distinct timeless approach involving 
{\it evolving constants of the motion} (`Heisenberg' rather than `Schr\"{o}dinger' style QM), and 
{\it partial observables} \cite{Rovellibook}. 
This is used e.g. in Loop Quantum Gravity's {\it master constraint program} \cite{Thiemann}).  

\noindent However, my main interest is in combining B) to D) (for which RPM's are well-suited), and 
which is a particularly interesting prospect \cite{H03} along the following lines.  
There is a records theory within histories theory.  
Histories decohereing is one possible way of obtaining a semiclassical regime in the first place.  
What the records are will answer the also-elusive question of which degrees of freedom are decohering 
which others in Quantum Cosmology.

%========================================================================================================
\subsection{Modelling assumptions and outline of the rest of this paper}
%========================================================================================================

In this paper I consider the case of 3 particles in 2-$d$ with scale ({\it scaled triangleland}), 
particularly at the quantum level.  
Scaled RPM is in some ways an easier theory than pure-shape RPM (it was found earlier, it is easier to 
reduce \cite{06I}). 
However, there are other ways in which it is not easier: maximal collision, the shape-scale 
interpretation of scaled RPM is an extension of the shape interpretation of pure-shape RPM.
This is relevant as regards solving the multi-HO potential problem in hand, for it involves 
scale--shape split form which is a natural extension of pure-shape RPM work in \cite{08II}.  
Thus I consider it {\sl after} the quantum treatment of pure-shape RPM \cite{08II}.  
Sec 2 covers kinematics. 
Firstly I describe triangleland in terms of relative Jacobi coordinates, and explain that the 
configuration space of pure shapes -- {\it shape space} -- of triangleland is $\mathbb{S}^2$.  
The {\it relational space}, that includes both the preceding and scale, is then the cone over 
$\mathbb{S}^2$.
This is $\mathbb{R}^3$, but with a non-flat metric; I exploit that it is, however, conformally flat. 
N.B. that the shape space is a nontrivial realization of $\mathbb{S}^2$, which involves Dragt-type 
\cite{Dragt} coordinates. 
I then interpret as shape quantities: ellipticity, anioscelesness and 
four times the area of the triangle.  
I finally provide a tessellation of the shape space sphere, that enables one to read off which points 
correspond to equilateral triangles, collinear configurations, double collisions, isosceles triangles 
etc, which is useful in interpreting dynamical trajectories, potentials and wavefunctions on shape space 
and relational space.

In Sec 3, I provide a classical prequel to the main quantum part of this paper, covering the 
`scale-dominates-shape' approximation and discussion of cosmologically-motivated choices of potential, 
as well as giving a qualitative account of the behaviour of the classical solutions.  
%
%Secs 2 and 3 represent applying the improvements of \cite{+tri, Cones} to the classical treatment 
%of scaled triangleland in \cite{08I}.  
%
\cite{AF, ScaleQM} considers the counterparts of this work for 4 particles on a line (which also has a 
$\mathbb{S}^2$ shape space, but here with a simple rather than `Dragt' realization).

Sec 4 sets up the quantum treatment of scaled triangleland, providing kinematical quantization, 
commutation relations, operator ordering, time-independent Schr\"{o}dinger equation and inner product 
in use. 
Sec 5 covers solutions of this with various soluble and/or cosmologically-motivated potentials.  
Much of this is analogous to, or an extension of, mathematics that occurs in the study of the atom.  
Some of this working is in spherical polar coordinates.
In the triangleland case, these respect the shape--scale split, with $I$ as radius and $\Theta$ a 
function of the ratio of relative separations of two constituent subsystems as azimuthal angle and 
the relative angle between these $\Phi$ as polar angle.  
Other parts of this working are in parabolic coordinates, which, for triangleland, are the same $\Phi$ 
as before and essentially the partial moments of inertia of the constituent subsystems.   
%
%This working uses some techniques developed at the classical level in \cite{08I}.
%
In Sec 6 I provide interpretation in terms of expectations, spreads and a Bohr moment of inertia for 
the model universe.  
This complements Sec 5's interpretation in terms of the probability density functions against the 
back-cloth of the tessellation of the sphere by its triangleland shape space interpretation.  
I also treat the inclusion of cosmological constant terms perturbatively.
The Conclusion (Sec 7) includes 1) some Problem of Time applications: hidden time, semiclassical 
time, histories theory and various timeless approaches.  
These are further developed in \cite{MGM, Forth}
2) Another application \cite{Forth} involves examples of further operator ordering and Dirac versus 
reduced quantization issues.

%========================================================================================================
%========================================================================================================
\section{Classical Kinematics}
%========================================================================================================
%========================================================================================================

%========================================================================================================
\subsection{Relative space and Jacobi coordinates}
%========================================================================================================

A configuration space for 3 particles in 2-$d$ is $\fQ = \mathbb{R}^{6}$.  
Rendering absolute position irrelevant (e.g. by passing from particle position coordinates to any sort 
of relative coordinates) leaves one on a configuration space {\it relative space} $\fR = \mathbb{R}^4$.  
The most convenient sort of coordinates for this are {\it relative Jacobi coordinates} \cite{Marchal} 
${\bf R}_1$ and ${\bf R}_2$.  
These are combinations of relative position vectors\footnote{I use 
%%%%%%%%%%%%%%%%%%%%%%%%%%%%%%%%%%%%%%%%%%%%%%%%%%%%%%%%%%%%%%%%%%%%%%%%%%%%%%%%%%%%%%%%%%%%%%%%%%%%%%%%%
$a$, $b$, $c$ as particle label indices running from 1 to $N$ for particle positions 
(usually $N$ = 3 in this paper).
$e$, $f$, $g$ as particle label indices running from 1 to $n = N - 1$ for relative position variables. 
These also run over the two parabolic coordinates as these are in 1 to 1 correspondence with 
the Jacobi relative position variables.
$i$, $j$, $k$ as spatial indices.
$p$, $q$, $r$ as relational space indices (in this paper's triangleland case, these run from 1 to 3).    
$u$, $v$, $w$ as shape space indices (in this paper's triangleland case, these run form 1 to 2).
[I reserve 
$d$ to denote dimension, 
$n$ for the number of relative position variables, 
$h$ for heavy and 
$l$ for light.]
I also use straight indices (upper or lower case) to denote quantum numbers, the index $S$ to denote 
`shape part' and the index $\rho$ (referring to the hyperradius) to denote `scale part'.}
%%%%%%%%%%%%%%%%%%%%%%%%%%%%%%%%%%%%%%%%%%%%%%%%%%%%%%%%%%%%%%%%%%%%%%%%%%%%%%%%%%%%%%%%%%%%%%%%%%%%%%%%%
${\bf r}^{ab} = {\bf q}^b - {\bf q}^a$ between particles into inter-particle cluster vectors that are 
such that the kinetic term is cast in diagonal form: ${\bf R}_1 = {\bf q}_3 - {\bf q}_2$ and ${\bf R}_2 
= {\bf q}_1 - (m_2{\bf q}_2 + m_3{\bf q}_3)/(m_2 + m_3)$.  
These have associated cluster masses $\mu_1 = m_2m_3/(m_2 + m_3)$ and $\mu_2 = m_1(m_2 + m_3)/(m_1 + 
m_2 + m_3)$.  
In fact, it is tidier to use one of the following. 
1) {\it mass-weighted} relative Jacobi coordinates ${\brho}^e = \sqrt{\mu_e}{\bf R}^e$ (Fig \ref{Fig1}). 
2) The squares of their magnitudes (partial moments of inertia) $I^e = \mu_e|{{\bf R}^e}|^2$. 
3) The normalized versions of 1), ${\bf n}^e := {\brho}^e/\rho$.   
Here, $\rho := \sqrt{I}$ and $I$ is the total moment of inertia. 
I use (a) as shorthand for \{a, bc\} where a,b,c form a cycle and a, bc are taken as a particular 
clustering (i.e. partition into subclusters).
I take clockwise and anticlockwise labelled triangles to be distinct.  
I.e. I make the plain rather than mirror-image-identified choice of set of shapes.
For specific components, I write the position indices downstairs as this substantially simplifies 
the notation.

%FFFFFFFFFFFFFFFFFFFFFFFFFFFFFFFFFFFFFFFFFFFFFFFFFFFFFFFFFFFFFFFFFFFFFFFFFFFFFFFFFFFFFFFFFFFFFFFFFFFFFFF
{            \begin{figure}[ht]
\centering
\includegraphics[width=0.15\textwidth]{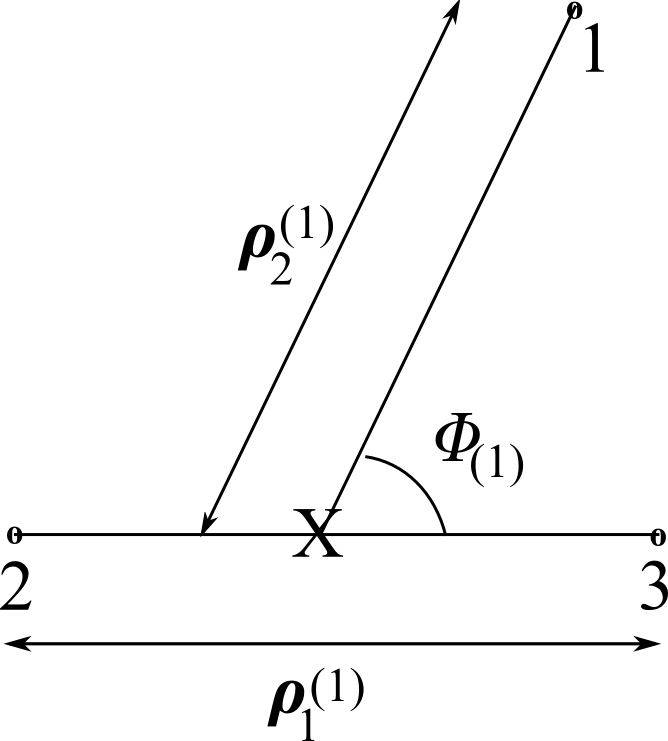}
\caption[Text der im Bilderverzeichnis auftaucht]{        \footnotesize{
\noindent
For 3 particles in the plane, one permutation of relative Jacobi coordinates are as indicated. 
X denotes the centre of mass of particles 2 and 3.  
In {\sl mass-weighted space}, the magnitudes are the length of a base of the triangle and what is a 
median in the equal-mass case. 
I define the `Swiss army knife' angle between the two $\brho_1^{(\sa)}$ by $\Phi_{(\sa)} = 
\mbox{arccos}\big({\brho}_1^{(\sa)}\cdot{\brho_2^{(\sa)}}/\rho_1^{(\sa)}\rho_2^{(\sa)}\big)$,  
and the ratio variable ${\Theta}_{(\sa)} = \mbox{arctan}(\rho_1^{(\sa)}/\rho_2^{(\sa)})$.    }         }
\label{Fig1}\end{figure}          }
%FFFFFFFFFFFFFFFFFFFFFFFFFFFFFFFFFFFFFFFFFFFFFFFFFFFFFFFFFFFFFFFFFFFFFFFFFFFFFFFFFFFFFFFFFFFFFFFFFFFFFFF

%========================================================================================================
\subsection{Relational space and shape space}
%========================================================================================================

If rotation with respect to absolute axes is to have no meaning, then one is left on a configuration 
space {\it relational space} ${\cal R} = \mathbb{R}^4/SO(2)$. 
If instead absolute scale were to have no meaning, then one is left on a configuration space 
\cite{Kendall} {\it preshape space} = $\mathbb{R}^4/\mbox{Dil}$ (for Dil the dilational group). 
It is straightforward to see that this is $\mathbb{S}^{3}$. 
If both of the above are to have no meaning, then one is left on \cite{Kendall} {\it shape space},  
$\fS = \mathbb{R}^4/SO(2) \times \mbox{Dil}$.   
Finally the relational configuration space is the cone over the shape space.  
At the topological level, for C(X) to be a cone over some topological manifold X, 
\beq
\mbox{C(X) = X $\times$ [0, $\infty$)/\mbox{ }$\widetilde{\mbox{ }}$} \mbox{ } , 
\eeq
where the meaning of $\widetilde{\mbox{ }}$ is that all points of the form (p $\in$ X, 0 $\in 
[0, \infty)$ ) are `squashed' i.e. identified to a single point termed the {\it cone point}, and 
denoted by 0. 
At the level of Riemannian geometry (see e.g. \cite{Mont98, ArchRat, Hsiang}), a cone C(X) over a 
Riemannian space X possesses a) the above topological structure and b) a Riemannian line element given 
by 
\beq
\d S^2 = \d{\cal R}^2 + {\cal R}^2\d s^2 \mbox{ } .
\eeq
Here, $\d s^2$ is the line element of X itself and ${\cal R}$ is a suitable `radial variable' that 
parametrizes the [0, $\infty$), which is the distance from the cone point.  
This metric is smooth everywhere except (possibly) at the troublesome cone point.

Now, $\mC(\mathbb{S}^2)$ is, at the topological level, $\mathbb{R}^3$.  
However, this $\mathbb{R}^3$ and $\mathbb{S}^2$ are not straightforward realizations at the level of 
configuration space metric geometry. 
The $\mathbb{R}^3$ has a curved metric on it and a dimensionally unintuitive radial variable, 
as follows.
The shape space sphere turns out to have radius 1/2, as can be seen from the relational space line 
element
\beq
\d S^2 = \d\rho^2 + (\rho^2/4)(\d\Theta^2 + \mbox{sin}^2\Theta\,\d\Phi^2)
\mbox{ } . 
\eeq
This inconvenience in coordinate ranges is then overcome by using $I$ instead as the radial variable, 
\beq
\d S^2 = (1/4I)(\d I^2 + I^2(\d\Theta^2 + \mbox{sin}^2\Theta\,\d\Phi^2)) 
\mbox{ } , \mbox{corresponding to } 
{\cal M}_{pq} = \mbox{diag}({1}/{4I}, \mbox{ } {I}/{4}, \mbox{ } {I\mbox{sin}^2\Theta}/{4}) 
\mbox{ } .
\eeq 
This metric ${\cal M}_{pq}$ is not the usual flat metric on $\mathbb{R}^3$: it is curved.  
However, it is clearly conformal to the flat metric
\beq
\d S^2_{\sf\sll\sa\st} = \d I^2 + I^2(\d\Theta^2 + \mbox{sin}^2\Theta\,\d\Phi^2) 
\mbox{ } , \mbox{corresponding to } 
\check{{\cal M}}_{pq} = \mbox{diag}
\left(
1, \mbox{ } I^2, \mbox{ } I^2\mbox{sin}^2\Theta
\right) \mbox{ } . 
\eeq
(This is in spherical polar coordinates with $I$ as radial variable, the conformal factor relating it to the 
previous metric being $\Omega^2 = 1/4 I$, a fact that is subsequently exploited in this paper). 
That $I$ features as radial variable is the start of significant differences between the triangleland 
and 4-stop metroland configuration spaces (the latter having the more intuitively obvious $\rho$ as 
radial variable).

Furthermore, for triangleland, the $\Theta$ and $\Phi$ are related to the corresponding clustering's 
${\bf n}^i$, and to relational space's unit Cartesian coordinates $u^i = (u_x, u_y, u_z)$, in a fairly 
unusual `Dragt-type' \cite{Dragt} way (closely related to the Hopf map):  
\beq
\mbox{dra}^{(\sa)}_x = \mbox{sin}\,\Theta_{(\sa)}\,\mbox{cos}\,\Phi_{(\sa)} = 
2\mn_1^{(\sa)}\mn_2^{(\sa)}\,\mbox{cos}\,\Phi_{(\sa)} = 2{\bf n}_1^{(\sa)}\cdot{\bf n}_2^{(\sa)}
\mbox{ } ,
\label{dragt1}
\eeq
\beq
\mbox{dra}^{(\sa)}_y = \mbox{sin}\,\Theta_{(\sa)}\,\mbox{sin}\,\Phi_{(\sa)} = 
2\mn_1^{(\sa)}\mn_2^{(\sa)}\,\mbox{sin}\,\Phi_{(\sa)} =  2({\bf n}_1^{(\sa)} \cr {\bf n}_2^{(\sa)})_3
\mbox{ } ,
\label{dragt2}
\eeq
\beq
\mbox{dra}^{(\sa)}_z = \mbox{cos}\,\Theta_{(\sa)} = \mn_2^{(\sa)\,2} - \mn_1^{(\sa)\,2} 
\mbox{ } .   
\label{dragt3}
\eeq  
Here, the 3-component in the second equation refers to the ficticious third dimension.

4-stop metroland, whose shape space is also $\mathbb{S}^2$, is also more straightforward in this 
respect, since its $u^i$ are just normalized relative separations $\rho^i$ \cite{+tri, Cones}. 
%
%for further comparison of the triangleland and 4-stop metroland configuration 
%spaces.  

%========================================================================================================
\subsection{Interpretation of Dragt coordinates in terms of scale and shape quantities}
%========================================================================================================

$\mbox{dra}^{(\sa)}_z$ is, by the difference form of (\ref{dragt3}), an ellipticity, ellip(a).  
This (and $\Theta_{(\sa)}$ itself) is a function of a pure ratio of relative separations.  
One can view dra$^{(\sa)}_x$ as a measure of `anisoscelesness' aniso(a), i.e. a departure (in a sense 
made precise in \cite{+tri}) from the notion of isoscelesness corresponding to the (a)-clustering.\footnote{C.f. 
%%%%%%%%%%%%%%%%%%%%%%%%%%%%%%%%%%%%%%%%%%%%%%%%%%%%%%%%%%%%%%%%%%%%%%%%%%%%%%%%%%%%%%%%%%%%%%%%%%%%%%%%%
anisotropy as a departure from isotropy in GR cosmology itself.}  
%%%%%%%%%%%%%%%%%%%%%%%%%%%%%%%%%%%%%%%%%%%%%%%%%%%%%%%%%%%%%%%%%%%%%%%%%%%%%%%%%%%%%%%%%%%%%%%%%%%%%%%%%
%
One can likewise view dra$^{(\sa)}_y$ as a measure of noncollinearity.   
Moreover this is actually a clustering-independent alias `democracy invariant' notion \cite{Zick, 
ACG86, LR95}.  
It is furthermore equal to four times the area of the triangle per unit $I$ in mass-weighted space.    
I denote $I$ by Size, to emphasize that it is a size variable, and I capitalize shape quantities 
to denote counterparts including a size factor: Dra$_{\Gamma}^{(\sa)} := 
I\, \mbox{dra}_{\Gamma}^{(\sa)}$, Ellip(a) $:= I$\, ellip (a), Aniso(a) $:= I$\, aniso(a), and 
Area $:= I$\, area, the actual area.

Thus there are 7 particular axes: 3 aniso(a), ellip(a) perpendicular pairs, all of which are 
perpendicular to an axis in the direction of the area vector.  
See Sec 2.5 for further study of the geometry involved.

%========================================================================================================
\subsection{Triangleland interpretation of parabolic coordinates}
%========================================================================================================

%FFFFFFFFFFFFFFFFFFFFFFFFFFFFFFFFFFFFFFFFFFFFFFFFFFFFFFFFFFFFFFFFFFFFFFFFFFFFFFFFFFFFFFFFFFFFFFFFFFFFFFFF 
{\begin{figure}[ht]
\centering
\includegraphics[width=0.67\textwidth]{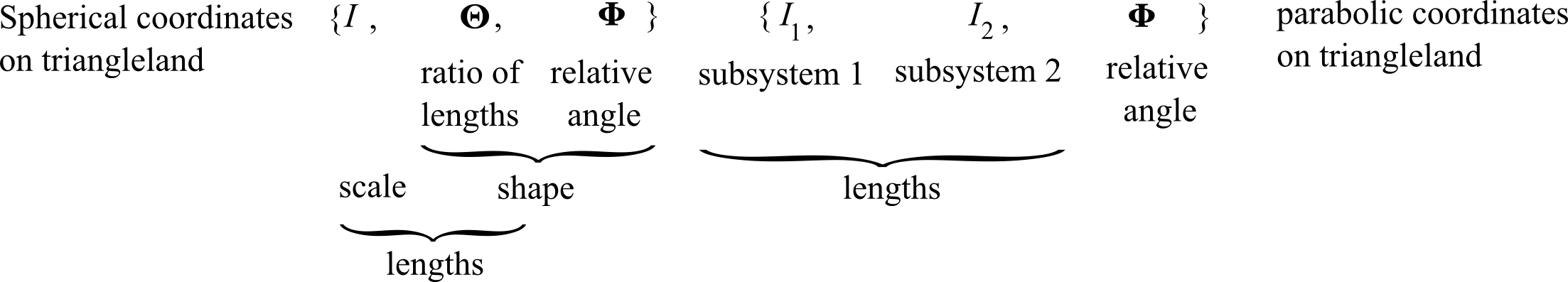}
\caption[Text der im Bilderverzeichnis auftaucht]{\footnotesize{Summary of the triangleland meanings of 
the spherical and parabolic coordinates used in this paper.}} \label{Fig-2}\end{figure} } 
%FFFFFFFFFFFFFFFFFFFFFFFFFFFFFFFFFFFFFFFFFFFFFFFFFFFFFFFFFFFFFFFFFFFFFFFFFFFFFFFFFFFFFFFFFFFFFFFFFFFFFFFF

\noindent 
Parabolic coordinates $\{\xi_1, \xi_2, \Phi\}$ on the conformally-related $\mathbb{R}^3$ are useful in 
this paper's quantum workings, much as they were already useful at the classical level in \cite{08I}.   
Here, $\Phi_{(\sa)}$ has the same meaning as before.  
The $\xi^{(\sa)}_{e}$, $e = 1, 2$ are  
\beq
\xi_1^{(\sa)} = 2I_1^{(\sa)} := 2 I \mbox{tall(\ma)} = I(1 - \mbox{ellip}(\ma)) 
%             = I(1 - \mbox{dra}_z^{(\sa)}) 
= I(1 - \mbox{cos}\Theta_{(\sa)}) 
\mbox{ } , \mbox{ } \mbox{ } 
\eeq
\beq
\xi_2^{(\sa)} = 2I_2^{(\sa)} := 2 I\,\mbox{flat(\ma)} = I(1 + \mbox{ellip}(\ma))  
%             = I(1 + \mbox{dra}_z^{(\sa)}) 
              = I(1 + \mbox{cos}\Theta_{(\sa)}) \mbox{ } .   
\eeq
Here, flat(a) = $\mn_1^{(\sa)\,2}$ is a flatness shape quantity (how much the clustering's base pair 
dominates the moment of inertia) whilst tall(a) = $\mn_2^{(\sa)\,2}$ is a tallness shape quantity (how 
much the particle not in the clustering's base pair dominates the moment of inertia of the system).   
%
%Then Tall(a) $:= I$\,tall(a) and Flat(a) $:= I$\,flat(a) *** check extent to which used ***.   
%
The flat configuration space metric is $\check{\cal M}_{pq}$ = 
diag$((\xi_1 + \xi_2)/4\xi_1, \mbox{ } (\xi_1 + \xi_2)/4\xi_2, \mbox{ } \xi_1\xi_2)$.

%========================================================================================================
\subsection{Tessellation of shape space and relational space}
%========================================================================================================

Assume equal masses for simplicity (see \cite{+tri} for elsewise).
Then distinguished points and curves on the triangleland shape space are as in Fig 3.
%
% LAYOUT: Hope that the figure pops up here in the final text! 
%
The use of this is that one can then interpret classical trajectories as paths upon this, and classical 
potentials and quantum-mechanical probability density functions as height functions over these.
Thus Fig 3 is useful as an `interpretational back-cloth' in subsequent Secs.

%FFFFFFFFFFFFFFFFFFFFFFFFFFFFFFFFFFFFFFFFFFFFFFFFFFFFFFFFFFFFFFFFFFFFFFFFFFFFFFFFFFFFFFFFFFFFFFFFFFFFFFFF 
{\begin{figure}[ht]
\centering
\includegraphics[width=0.7\textwidth]{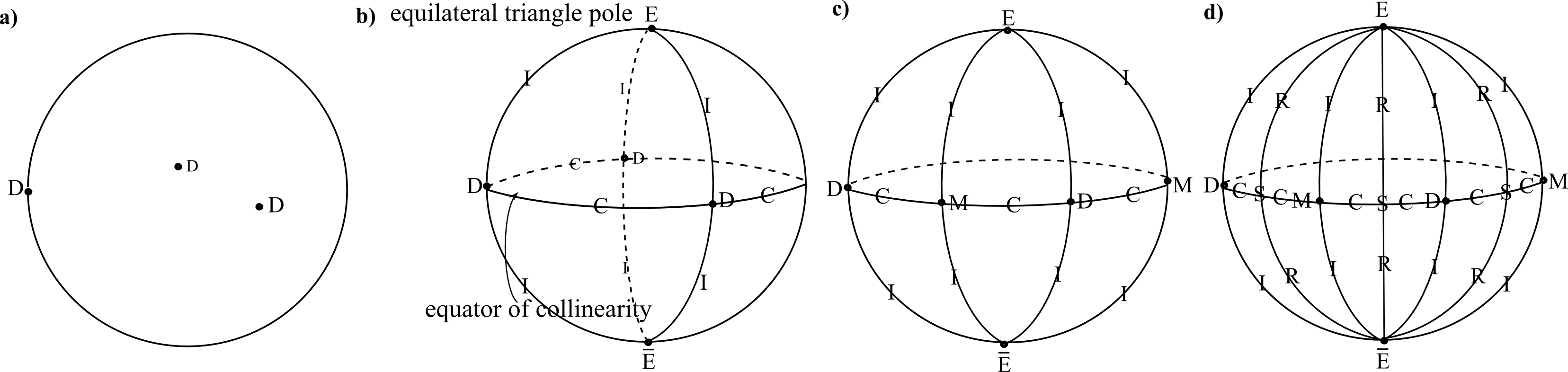}
\caption[Text der im Bilderverzeichnis auftaucht]{\footnotesize{ Tessellation of triangleland.  
\noindent a) Triangleland's shape space's topologically-relevant features, i.e. those depending 
solely on the notion of coincidence: three points corresponding to the possible permutations of double 
collision, D.  
If these points are excised, the resulting topology is the `pair of pants' \cite{ArchRat, Montgomery2}.

\noindent b) Next, triangleland's shape space's further notions of distance and angle at the level of 
metric geometry furbish 1) a notion of collinearity C that comprise the equator (split into 3 equal 
arcs by the D's).
This equator of collinearity is orientationless and separates the 2 hemispheres of distinct orientation.   
2) A notion of equilaterality E (and its labelled mirror image $\bar{\mE}$.) which occur at the poles.  
3) The half great circles joining the D's to the E's correspond to the 3 distinct notions of isosceles 
triangles, I, that a labelled triangle possesses. 
Each notion of isoscelesness is associated with a particular clustering (the one that picks out the 
particular base about the bisector of which one has mirror symmetry about).
On the other hand, the notions of collinearity and equilateralness are clearly clustering-independent.
They are the plane of zero area and points of extremal area per given $I$, i.e. the E-direction 
perpendicular to the collinearity plane is proportional to the area vector.  

\noindent c) However, it is not just half, but the whole of, the abovementioned great circles that 
correspond to isosceles triangles.  
There are planes of zero anisoscelesness, the perpendiculars 
to each of which are proportional to the corresponding aniso(a) vector.      
This plane separates hemispheres of Left(a) [left-slanting of (a)'s notion of isoscelesness] and 
Right(a) [right-slanting of (a)'s notion of isoscelesness].  
That b) only captured half of each isoscelesness great circle I also attracts attention to the other  
intersection points of this and the collinearity equator C (which are antipodal to the D's).  
These are {\it mergers}, M: configurations such that one particle is at the centre of mass of the other 
two.  
This is clearly another clustering-dependent notion.  

\noindent d) In wanting to form axis systems, the great circles perpendicular to the I's are also 
of interest.  
These are the {\it regular configurations}, R (given by $I_1^{(\sa)} = I_2^{(\sa)}$).
These are planes of zero ellipticity, the perpendiculars to each of which are proportional to the 
corresponding ellip(a) vector.
These each separate two hemispheres of clustering-dependent notions of tall(a) 
($I_2^{(\sa)} > I_1^{(\sa)}$) and flat(a) ($I_1^{(\sa)} > I_2^{(\sa)}$) triangles. 
I then know of no further interesting feature possessed by the intersection points of C and R, so I 
term these `spurious' points, S.  

%\noindent e) This depicts how the plain triangleland contains the version of 3-stop metroland that is 
%mirror-image-identified (due to the availability of the second dimension to rotate the one image into 
%the other.)  
} } \label{Fig-3tri}\end{figure}            } 
%FFFFFFFFFFFFFFFFFFFFFFFFFFFFFFFFFFFFFFFFFFFFFFFFFFFFFFFFFFFFFFFFFFFFFFFFFFFFFFFFFFFFFFFFFFFFFFFFFFFFFFFF

\noindent The underlying symmetry group is the order-12 $\mathbb{D}_3 \times \mathbb{Z}_2 \mbox{ } 
\widetilde{=} \mbox{ } S_3 \times \mathbb{Z}_2$, so that it corresponds to the freedom of relabelling 
the 3 partices and of ascribing an overall orientation.  
%
%The tessellations in Figs 3c) and d) then have larger symmetry groups, $\mathbb{D}_{6} \times 
%\mathbb{Z}_2$  of order 24 and $\mathbb{D}_{12} \times \mathbb{Z}_2$  of order 48, but once the labels 
%are taken into account, one returns to $\mathbb{D}_{3} \times \mathbb{Z}_2$. 
%
Also, note that the D and M vertices pick out 3 particularly distinguished axes at $\pi/3$ to each 
other in the collinearity plane. 
There being 3 of these corresponds to there being 3 permutations of Jacobi coordinates, with each such 
axis corresponds to an ellip(a).   
The axes perpendicular to each of these correspond to the 3 notions of aniso(a).  
I denote spherical polar coordinates about each of these as principal axis by \{$\Theta_{(\sa)}$, 
$\Phi_{(\sa)}$\}, where $\Theta_{(\sa)}$ is the azimuthal spherical angle as measured from each D. 
I denote the more `natural' spherical polar coordinates with E as North pole and the ath DM as second 
axis by $\{I, \Theta_{[\sa]}, \Phi_{[\sa]}\}$.

Triangleland's relationalspace is the corresponding cone over the above tessellation-decorated shape 
space.
E, D, M, S become half-lines, C, I, R become sectors of 2-$d$ angle, and faces become sectors of solid 
angle, all emanating from the triple collision at the cone point, 0.

% LAYOUT: 5 pages is a good mark for the end of kinematics.  

%========================================================================================================
%========================================================================================================
\section{Classical dynamics of triangleland}
%========================================================================================================
%========================================================================================================

%========================================================================================================
\subsection{Shape-scale split action on relational space}
%========================================================================================================

One form for the relational action for scaled RPM is\footnote{For relational space, ${\cal Q}^{p}$, 
%%%%%%%%%%%%%%%%%%%%%%%%%%%%%%%%%%%%%%%%%%%%%%%%%%%%%%%%%%%%%%%%%%%%%%%%%%%%%%%%%%%%%%%%%%%%%%%%%%%%%%%%% 
${\cal P}_{p}$, ${\cal M}_{pq}$, ${\cal M}$ and ${\cal N}^{pq}$ are general coordinates, 
their conjugate momenta, the metric, its determinant and its inverse.  
For shape space, I denote the counterparts of these by $S^{u}$, $P_{u}$, 
$M_{uv}$, ${M}$ and ${N}^{uv}$.}  
%%%%%%%%%%%%%%%%%%%%%%%%%%%%%%%%%%%%%%%%%%%%%%%%%%%%%%%%%%%%%%%%%%%%%%%%%%%%%%%%%%%%%%%%%%%%%%%%%%%%%%%%%
\beq
\fS = 2\int\sqrt{\check{T}(\check{E} - \check{V})}\d\lambda \mbox{ } , 
\mbox{ } 
\label{accio}
\eeq
where $\lambda$ is a label time and $\mbox{}^{\prime} = \d/\d\lambda$.
$\check{T} = \check{\cal M}_{pq}{\cal Q}^{p\prime}{\cal Q}^{q\prime}/2$ is the kinetic term, which 
splits into scale and shape parts
\beq
\check{T} = T_{\rho} + \rho^2\fT_{S} \mbox{ } , \mbox{ } \mbox{ } 
T_\rho = {\rho}^{\prime\,2}/2 \mbox{ } , \mbox{ } \mbox{ } 
\fT_{S} = M_{uv}S^{u\prime}S^{v\prime}/2 = 
({\Theta}^{\prime\,2} + \mbox{sin}^2\Theta\,{\Phi}^{\prime\,2})/2 
\eeq

\noindent(the last equality being for triangleland in spherical coordinates).  
This action is `banal conformal invariant' under $T \longrightarrow \Omega^2 T$, 
$E - V \longrightarrow (E - V)/\Omega^2$.  
In displaying the above form, one has already moved from the mechanically natural version to the 
banal-conformally related flat metric version that I denote by checking, which corresponds to 
$\Omega^2 = 1/4I$ (so that $\check{V} = V/4I$ and $\check{E} = E/4I$).

The momentum--velocity relations and the equations of motion then feature $\sqrt{(\check{E} - \check{V})
/\check{T}}\,\mbox{ }^{\prime}{\mbox{ }} := \d/\d \check{t}^{\se\sm} := \check{\dot{\mbox{ }}}$, for 
$\check{t}^{\se\sm}$ one of the relational approach's banally-related {\sl emergent time} notions.  
The unchecked version of this is a recovery of Newtonian, proper and cosmic time 
%
%This last one could be false... which could be rather interesting.  
%
in various different contexts, as well as coinciding with the semiclassical notion of emergent time 
mentioned around eq (\ref{HamJac}).  
It is then this {\sl joint} notion of emergent time that I denote by $t^{\se\sm}$. 
The equations of motion following from this action are provided in e.g. \cite{Cones}.

I remark here that as well as $E$ being conserved, one has a conserved relative angular momentum 
${\cal J}$ for $V$ independent of $\Phi$.
If $V$ is independent of $\Theta$ also, one has, rather, three conserved quantities 
${\cal R}\mbox{at}_{\Delta}$, standing for relative {\it rational} momenta (a generalization of angular 
momenta covering both angular momenta and dilational momenta contributions \cite{AF,+tri,Cones}).
[Here, I let capital Greek indices run over the $SO(3)$ generators ${\cal J} = {\cal R}\mbox{at}_3$.] 
I then use ${\cal T}\!\mo\mt$ to denote $\sum_{\Delta = 1}^3{\cal R}\mbox{at}_{\Delta}^2$ i.e. total 
rational momentum.

%========================================================================================================
\subsection{Choice of potentials}
%========================================================================================================

Pure-shape RPM has much potential freedom (any potential homogeneous of degree 0 is valid), 
while scaled RPM can have any potential at all.
So far, the RPM program has studied free problems and HO potentials (or HO-like ones for 
pure-shape RPM, for which HO potentials themselves are disallowed by the homogeneity requirement).   
These are $\rho^2(A + B\,\mbox{cos}\,2\varphi)$ for 3-stop metroland and $\rho^2(A + 
B\,\mbox{cos}\,2\theta + C\,\mbox{sin}^2\theta\,\mbox{cos}\,2\phi)$ for 4-stop metroland.  
%
%and $I^2(A + B\,\mbox{cos}\,2\Theta + C\,\mbox{sin}^2\Theta\,\mbox{cos}\,2\Phi )$ for triangleland 
%in the flat banal representation, which is also accompanied by a $E/4I$ term.  
%
Advantages of such potentials are boundedness and good analytical tractability.  
However, we do need to look at further models due to their atypical simplicity and only 
partially-relevant parallel the dominant scale dynamics of commonly-used cosmological models.

The shape-scale split also holds approximately if there was shape dependence but it and 
changes in it are small compared to changes of scale.  
This requires some stability condition so as to be applicable long-term. 
In this case one has $V_{(0)}$ in place of $V$.   
Obtaining a separated-out heavy slow scale part for the semiclassical approach relies on the following 
`scale-dominates-shape' approximation being meaningful,

\beq
|T_{l}| << |T_{h}| \mbox{ realized by the shape quantity } \rho^2\fT_{S} <<   T_{\rho} 
\mbox{  (scale quantity) } ,
\label{SSA1}
\eeq
\beq
|J_{hl}| << |V_{h}| \mbox{ realized by } |J(\rho, S^{u})| << |V_{(0)}(\rho)|  
\mbox{ for }
\label{SSA2}
\eeq
for $J_{hl}$ the interaction part of the potential.  
\beq
V(\rho, S^{u}) \approx V_{(0)}(\rho)(1 + V_{(1)u}(\rho)S^{u} + 
O(|S^{u}|^2) = V_0(\rho) + J(\rho, S^{u}) \mbox{ } .  
\label{SSA3}
\eeq 
[Without such approximations, one cannot separate out the heavy (here scale) part so that it can 
provide the approximate timefunction with respect to which the light (here shape) part's dynamics runs.]
Counterparts of this in GR Quantum Cosmology are leading-order neglect of scalar field terms  
\cite{GibGri}, of anisotopy \cite{Amsterdamski} and of inhomogeneity \cite{HallHaw}.  

%\mbox{ } 

Next, I use the analogy between Mechanics and Cosmology to broaden the range of potentials under 
consideration and pinpoint ones which parallel 
classical and Quantum Cosmology well. 
This would seem to be making more profitable use of the potential freedom than previous mere use of 
simplicity.
Isotropic cosmology (in $c = 1$ units) has the Friedmann equation
\beq
\left(\frac{\dot{a}}{a}\right)^2 = - \frac{k}{a^2} + \frac{8\pi G\epsilon}{3} + \frac{\Lambda}{3}  
= -\frac{k}{a^2} + \frac{2GM_{\sd\su\sss\st}}{a^3} + \frac{2GM_{\sr\sa\sd}}{a^4} + \frac{\Lambda}{3}
\mbox{ } ,    
\eeq
the second equality coming after use of energy--momentum conservation and assuming noninteracting matter 
components. 
Here, $a$ is the scalefactor of the universe, $\dot{\mbox{ }} = \d/\d t^{\sc\so\sss\sm\si\sc}$ (which is GR's $\d/\d t^{\se\sm}$ 
here). 
$k$ is the spatial curvature which is without loss of generality normalizable to 1, 0 or --1. 
$G$ is the gravitational constant, $\epsilon$ is matter energy density.
$\Lambda$ is the cosmological constant. 
$M$ is the mass of that matter type that is enclosed up to the radius $a(t)$. 
A fairly common analogy is then between this and (unit-mass ordinary mechanics energy equation)$/r^2$, 
\beq
\left(\frac{\dot{r}}{r}\right)^2 = \frac{2E}{r^2} + \frac{K_{\mbox{\scriptsize Newton}}}{r^3} + 
\frac{K_{\mbox{\scriptsize Conformal}}}{r^4} + K_{\mbox{\scriptsize Hooke}}
\mbox{ }    
\eeq
where here and elsewhere in this paper the various $K$'s are constant coefficients, and 
$\dot{\mbox{ }} = \d/\d t^{\se\sm}$ is now also identified as $\d/d t^{\sN\se\sw\st\so\sn}$.  
A particularly well-known subcase of this is that 1-$d$ mechanics with a $1/r$ Newtonian gravity type 
potential is analogous to isotropic GR cosmology of dust. 
This extends to an analogy between the Newtonian dynamics of a large dust cloud and the GR isotropic 
dust cosmology \cite{CosMech,BGS03}.
Here, shape is least approximately negligible through its overall averaging 
out to approximately separated out shape and cosmology-like scale problems.  
For this paper's purposes, enough of these parallels (\cite{HarMTWPee, Rindler}) 
survive  1)the introduction of a pressure term in the cosmological part. 
2) the using of the spherical presentation of triangleland RPM in place of ordinary mechanics.

It is between the above Friedmann equation and (energy relation)/$I^2$,
\beq
\left(\frac{\dot{I}}{I}\right)^2 =  \frac{2E}{I^3} - \frac{{\cal T}\!\so\st}{I^4} 
                                                     - \frac{2V(I, \, S^u)}{I^3} 
= \frac{-2A}{I^2} + \frac{2E}{I^3} + \frac{2R - {\cal T}\!\so\st}{I^4} - 2\sigma
\mbox{ } . 
\eeq 
Thus the spatial curvature term $k$ becomes the net HO terms' $2A$.   
The cosmological constant term $\Lambda/3$ becomes --2 times the surviving lead term from the 
$r^{IJ \, 6}$ potentials's coefficient, $\sigma$.  
The dust term $2GM/a^3$'s coefficient $2GM$ becomes $2E$, so for analogy with a physical dust, we need 
$E > 0$.
The radiation term coefficient $2GM$ becomes $- {\cal T}\!\mo\mt + 2R$ for $2R$ the coefficient of the 
$V_{(0)}$ contribution from the $1/r^{ab\,2}$ terms.  
This is again the conformally invariant potential term.

Note that ${\cal T}\!\mo\mt$ itself is of the wrong sign to match up with the ordinary radiation term 
of Cosmology.  
In the cosmological GR context, `wrong sign' radiation fluid means that it still has $p = \epsilon/3$ 
equation of state (for $p$ the pressure), but its density $\epsilon$ is negative.  
This violates all energy conditions, making it unphysical in a straightforward GR cosmology context, 
and also has the effect of singularity theorem evasion by `bouncing'.  
But from the ordinary mechanics perspective, this is just the well-known repulsion of the 
centrifugal barrier that prevents collapse to zero size.
This difference in sign is underlied by an important limitation in Mechanics--Cosmology analogies. 
Namely, that in mechanics, the kinetic energy is positive-definite, while in GR the kinetic energy is 
indefinite, the scale part contributing negatively.  
This does not affect most of the analogy because the energy and potential coefficients can be defined 
with opposite signs. 
However, the relative sign of the shape and scale kinetic terms cannot be changed by such a manoeuvre, 
and it is from this that what is cosmologically the `wrong sign' arises. 
Two different arguments about what to do with the wrong-sign term are   
1) one can suppress this ${\cal T}\!\mo\mt$ term that signifies the relative rational momentum of two 
constituent subsystems.
One can do so by taking toy models in which it is zero, or small, or swamped by `right-sign' 
$1/|{\bf r}^{ab}|^2$ contributions to the potential. 
2) More exotic geometrically complicated scenarios such as brane cosmology can possess `dark radiation' 
including of the `wrong sign' \cite{DarkRad}. 
(These can indeed possess what appears to be energy condition violation from the 4-$d$ spacetime 
perspective due to projections of higher-dimensional objects \cite{AT05}).  
Thus such a term is not necessarily unphysical.

In the spherical triangleland analogy, the Newtonian $1/|{\bf r}^{ab}|$ type potentials that one might 
consider to be mechanically desirable to include produce $1/I^{7/2}$ terms. 
These are analogous to an effective fluid with equation of state $P = \epsilon/6$ (for $P$ the pressure) 
i.e. an interpolation `halfway between' radiation fluid and dust. 
This is physically reasonable for a cosmology: it does not violate any energy conditions. 
It is sensible as a rough model of a mixture of dust and radiation as is believed to have been present 
when the universe was around 60000 years old.

\noindent The spherical presentation of triangeland's analogy with Cosmology is different from 
the complex projective space presentation's (\cite{Cones} contrasts these two different Cosmology--RPM 
analogies).  
The former is of limited use due to not extending to higher `$N$-a-gonlands'. 
However, one use for it is that it allows the shape part to be studied in $\mathbb{S}^2$ terms which 
more closely parallel the Halliwell--Hawking \cite{HallHaw} analysis of GR inhomogeneities over 
$\mathbb{S}^3$.

What the Cosmology--RPM mechanics analogy provides at the semiclassical level is a slow heavy scale 
dynamics paralleling that of Cosmology. 
Moreover, it is now coupled to a light fast shape dynamics that is simpler than GR's while retaining a 
meaningful notion of locality/inhomogeneity/structure.
The analogy does not however go as far as having a metric interpretation or a meaningful interpretation 
in terms of an energy density $\epsilon$; triangleland is, after all, just a particle mechanics model, 
and thus has no such notions.

%=======================================================================================================
\subsection{Hydrogen analogy for triangleland with very special multi-HO potential}
%=======================================================================================================

The very special scaled RPM HO banal-conformally maps to the hydrogenic/Newtonian gravity problem with 
\beq
\mbox{ (radius) } =  r \mbox{ } \longleftrightarrow \mbox{ } I \mbox{ (total moment of inertia) } , 
\mbox{ } 
\eeq
\beq
\mbox{ (test mass) } =  m \mbox{ } \longleftrightarrow \mbox{ } 1 \mbox{ } , \mbox{ } 
\eeq
\beq
\mbox{ (angular momentum) }   =       \mL \mbox{ } \longleftrightarrow \mbox{ } J \mbox{ (relative 
angular momentum -- see App C) } , \mbox{ } 
\eeq
\beq
\mbox{ (total energy) } = {\cal E} \mbox{ } \longleftrightarrow \mbox{ } - A = 
\mbox{ -- (sum of mass-weighted Jacobi--Hooke coefficients)/16 }   
\eeq
and the 1-electron Coulomb problem
\beq
\mbox{ (nuclear charge)(test charge of electron)/4$\pi$(permettivity of free space) } = 
(Ze)e/4\pi\epsilon_0 \mbox{ } \longleftrightarrow E  \mbox{ (total energy)/4} 
\eeq
[or to the Newtonian gravitation problem with the last analogy replaced by 
\beq
\mbox{ (Newton's gravitational constant)(massive mass)(test mass) } = GMm \mbox{ } \longleftrightarrow 
\mbox{ } E \mbox{ (total energy)/4 }] \mbox{ } .    
\eeq
Note that the positivity of the Hooke's coefficients translates to the requirement that the 
gravitational or atomic energy be negative, i.e. to bound states.  
Also note that the positivity of $E$ required for classical consistency corresponds to attractive 
problems like the Kepler or atomic problem being picked out, as opposed to repulsive Coulomb problems.

%========================================================================================================
\subsection{Outline of the behaviour of approximately-classical solutions}
%========================================================================================================

While the Cosmology--RPM analogy map is different in each of this paper and \cite{ScaleQM}, 
in each case it maps to the same set of fairly well-known cosmologies. 
Thus there is no need for an separate case-by-case classical qualitative study to that in 
\cite{ScaleQM}.  
[This is no longer true at the quantum level, however, due to the differences in inner product 
between scaled triangleland and scaled 4-stop metroland.]

The issue of stability concerns what potentials are needed to approximately map to the well-known 
cosmologies.  

\noindent
\cite{ScaleQM} had some problems here due to negative power law potentials being required, 
for which the scale-dominates-shape approximation broke down.  
However, the difference in the analogy between Cosmology and the spherical presentation of triangleland 
causes the potentials studied in this paper to be purely positive powers for which no 
such problem occurs.

Forms of the approximate heavy-$I$ solutions for further use in the Semiclassical approach part of 
the conclusion are now as follows.  
$I =  \mbox{sin}(\sqrt{2S}t^{\se\sm})/\sqrt{2S}$ is analogous to the Milne in anti de Sitter 
solution \cite{Rindler}.
$I =  \mbox{cosh}(\sqrt{-2S}t^{\se\sm})/\sqrt{-2S}$ is analogous to the positively-curved 
de Sitter model. 
[Both are models with just analogues of $k$, $\Lambda$ of various signs.]
$I = (9E/2)^{1/3}t^{\se\sm\,2/3}$ is analogous to the flat dust model, with well known 
cycloid and hyperbolic counterpart solutions in the positively and negatively-curved cases.  
$I = (4(2R - {\cal T}\!\mo\mt))^{1/4}t^{\se\sm \, 1/2}$ is analogous to 
the flat radiation model, with well-known curved counterparts (one analogous to the Tolman model). 
The present problem also requires cosmologically less familiar wrong-sign radiation models.   
However, these are familiar in ordinary mechanics as solutions with central term, 
such as $I = \sqrt{t^{\se\sm \, 2} + {\cal T}\!\mo\mt}$.

%======================================================================================================== 
%=========================================================================================================
\section{Quantum treatment}
%========================================================================================================
%========================================================================================================

%========================================================================================================
\subsection{Kinematical quantization}
%========================================================================================================

An appropriate kinematical quantization \cite{Isham84} for scaled triangleland involves $\mathbb{R}^3 
\mbox{\textcircled{S}} \mbox{Eucl}(3)$. 
Here, Eucl(3) is the Euclidean group of translations and rotations, Tr(3) $\mbox{\textcircled{S}}$ 
Rot(3) = $\mathbb{R}^3 \mbox{\textcircled{S}} SO(3)$.  
$\mbox{\textcircled{S}}$ denotes a semi-direct product.  
A good kinematical choice of objects is 
then Dra$^{\Gamma}$ for the first $\mathbb{R}^3$, the translational generators 
$\Pi^{\sD\sr\sa}_{\Gamma}$ and the $SO$(3) generators ${\cal R}\ma\mt_{\Gamma}$.  
Then the nontrivial commutation relations between these are 
\beq
[\widehat{\mR\ma\mt}_{\Gamma}, \widehat{\mR\ma\mt}_{\Delta}] = 
i\hbar\epsilon_{\Gamma\Delta}\mbox{}^{\Lambda}\widehat{\mR\ma\mt}_{\Lambda} 
\mbox{ } , \mbox{ } \mbox{ } 
[\widehat{\mD\mr\ma}\mbox{}^{\Gamma}, {\mR\ma\mt}_{\Delta}] = 
i\hbar\epsilon^{\Gamma}\mbox{}_{\Delta}\mbox{}^{\Lambda}\widehat{\mD\mr\ma}\mbox{}^{\Lambda} 
\mbox{ } , \mbox{ } \mbox{ } 
[\widehat{\Pi}^{\sD\sr\sa}_{\Gamma}, {\mR\ma\mt}_{\Delta}] = i\hbar\epsilon_{\Gamma\Delta}\mbox{}^{\Lambda}
\widehat{\Pi}_{\Lambda}^{\sD\sr\sa}  \mbox{ } .
\eeq

%========================================================================================================
\subsection{Operator-ordering and time-independent Schr\"{o}dinger equations}
%========================================================================================================

The {\it Laplacian ordering} at the QM level of the classical combination 
${\cal N}^{pq}({\cal Q}^{r}){\cal P}_{p}{\cal P}_{q}$ in the quadratic constraint is 
\beq
D^2 = \frac{1}{\sqrt{\cal M}} \frac{\pa}{\pa {\cal Q}^{p}}
\left(
\sqrt{\cal M}{\cal N}^{pq}\frac{\pa}{\pa{\cal Q}^{q}}
\right) \mbox{ } . 
\eeq  
This has the desirable property of (straightforwardly) being independent of coordinate choice on the 
configuration space $\fQ$ \cite{DeWitt57}. 
However, this property is not unique to this ordering; one can reorder to include a Ricci scalar 
curvature term so as to have $D^2 - \xi\,\mbox{Ric}({\cal M})$ \cite{DeWitt57, Oporder, Magic, HP86}. 
Among these, there is a unique choice of $\xi$ (dependent on the dimension $k \geq 2$ of the 
configuration space) which gives a conformally-invariant operator-ordering \cite{Oporder, HP86}: 
\beq
{D}_{\sc}^2 = \frac{1}{\sqrt{\cal M}} \frac{\pa}{\pa {\cal Q}^{p}}
\left(
\sqrt{\cal M}{\cal N}^{pq}\frac{\pa}{\pa{\cal Q}^{q}}
\right) 
- \frac{k - 2}{4(k - 1)}\mbox{Ric}(M) \mbox{ }  .  
\eeq
Moreover, in \cite{Banal} I identified this conformal invariance to be the same as the banal conformal 
invariance that the relational product-type action manifests (c.f. Sec 3.1).  
Also note that operator (33) is by itself is still not banal-conformally invariant.  
For, it is furthermore required that the wavefunction of the universe $\Psi$ that it acts upon itself 
transforms in general tensorially under banal conformal transformations (paralleling \cite{Wald}), 
\beq
\Psi \longrightarrow \widetilde{\Psi} = \Omega^{\frac{2 - k}{2}}\Psi \mbox{ } .
\eeq

In the specific case of this paper, I work in the checked presentation.  
Then Ric($\check{\cal M}$) = 0, so the conformal ordering just collapses to the Laplacian one, and this 
Laplacian is just the obvious analogue of the usual spherical one: 
\beq
D^2_{\sc} = D^2 = \frac{1}{I^2}
\left(
\frac{\pa}{\pa I}
\left( 
I^2\frac{\pa}{\pa I} 
\right) + D^2_{\mathbb{S}^2}
\right) 
\mbox{ } .
\eeq
Thus the time-independent Schr\"{o}dinger equation is
\beq
\frac{-\hbar^2}{2I^2}
\left( 
\frac{\pa}{\pa I}
\left( 
I^2\frac{\pa}{\pa I} 
\right) 
+ \frac{1}{I^2}D^2_{\mathbb{S}^2}
\right)
\check{\Psi} + \frac{V}{4I}\check{\Psi} = \frac{E}{4I}\check{\Psi} \mbox{ } .  
\eeq
If $V/4I = \check{V}_I + \check{V}_{\mathbb{S}^2}$, 
$\Psi(I, \Theta, \Phi) = {F}(I){\cal S}(\Theta, \Phi)$
separates this into 
$
-\hbar^2D_{\mathbb{S}^2}{\cal S} + \check{V}_{\mathbb{S}^2}{\cal S} = \lambda{\cal S}$,
which, for $\check{V}_{\mathbb{S}^2} = 0$ case common in this paper, gives ${\cal S} = Y_{\sR\sr}
(\Theta, \Phi)$ spherical harmonics and $\lambda = \hbar^2\mR(\mR + 1)/2$, $\mR \in \mathbb{N}$ and 
then the separated-out scale equation is   
\beq
-\frac{\hbar^2}{2}
\left(
\frac{1}{I^2}\frac{\pa}{\pa I}
\left( 
I \frac{\pa{F}}{\pa I}  
\right) 
- \frac{   \mR(\mR + 1){F}    }{    I^2    }
\right) 
+ \check{V}_{I}{F} = \frac{E}{4I}{F} \mbox{ } .
\eeq
In the more general case of $\Phi$-independent $V$, I use r for the projected rational 
momentum quantum number.

Also useful below, the parabolic coordinates version of the time-independent Schr\"{o}dinger equation is

\noindent
\beq
-\frac{\hbar^2}{2}
\left(
\frac{4}{\xi_1 + \xi_2}
\left(
\frac{\pa}{\pa\xi_1}
\left( 
\xi_1\frac{\pa}{\pa\xi_1}
\right)
+ \frac{\pa}{\pa\xi_2}
\left( 
\xi_2\frac{\pa}{\pa\xi_2}
\right)
\right)
+ 
\left(
\frac{1}{\xi_1} + \frac{1}{\xi_2}
\right)
\frac{\pa^2}{\pa\Phi^2}
\right)
\check{\Psi} + \frac{V\check{\Psi}}{2} = \frac{E\check{\Psi}}{2} \mbox{ } ,  
\eeq
which is separable if $V = V_1(\xi_1) + V_2(\xi_2)$ alone, via 
$\check{\Psi}(\xi_1, \xi_2, \Phi) = \phi(\Phi)\Xi_1(\xi_1)\Xi_2(\xi_2)$ into: 
simple harmonic motion and 
\beq
-4\hbar^2\frac{\pa}{\pa\xi_{e}}
\left(
\xi_{e}\frac{\pa\Xi_{e}}{\pa\xi_{e}} + \frac{\hbar^2j^2\Xi_{e}}{\xi_{e}}
\right) 
+ V_{e}(\xi_{e})\Xi_{e} = E_{e}\Xi_{e} \mbox{ } . \label{xiano}
\eeq
for $E_{e}$, $e = 1, 2$ separation constants such that $E_1 + E_2 = E$.

%========================================================================================================
\subsection{Inner product in use}
%========================================================================================================

In passing between the plain and checked banal representations, the inner product is mapped according to 
\cite{Banal}
\beq
\int \Psi_1^*\Psi_2 \sqrt{M}\d^3x \mbox{ } \longrightarrow \mbox{ } 
\int \check{\Psi}_1^*\check{\Psi}_2 \Omega^{-2}\sqrt{\check{M}}\d^3x
\eeq
where the particular confomal factor is $\Omega^2 = 4I$.
Now, in the checked representation, $\sqrt{\check{M}}$ is the usual spherical Jacobian 
$I^2\mbox{sin}\Theta$, so the checked inner product weight is $I\,\mbox{sin}\Theta\, /4$. 
Alternatively and equivalently, the plain inner product is $(1/4I)^{3/2}I^2\mbox{sin}\Theta = 
I^{1/2}\mbox{sin}\Theta\, /8$, 
i.e. differing from the usual spherical one by the obvious conformal factor. 
Also, the checked representation in parabolic coordinates, the inner product weight is just $1/8$.

Conformal-transforming prior to solving parallels that which is done by e.g. Iwai, Tachibana and Uwano 
\cite{IwaiUwano} for a different problem (the 4-$d$ isotropic HO). 
As they explain well, solving QM involves wavefunctions {\sl and} inner product being found, in which 
case this paper does {\sl not} involve hydrogen.  
For, it has been set up to have the same wavefunctions as hydrogen but the inner products are different 
(so e.g. normalization is different, as are expectations of operators).

%========================================================================================================
%========================================================================================================
\subsection{No Monopole issues}
%========================================================================================================
%========================================================================================================

\cite{Cones} establishes that the present study of relational triangleland involves no monopole effects 
in the study of triangleland, due to $L = 0$ being an analogous simplification to particles carrying 
no charge.

%========================================================================================================
%========================================================================================================
\section{QM solutions}
%========================================================================================================
%========================================================================================================

%========================================================================================================
\subsection{Useful bases for triangleland}\label{Bas}
%========================================================================================================

Looking at the tessellation, there is one particularly distinguished choice for principal axis: E. 
Then a distinguished choice for second axis is D, with the third one then being toward the R 
perpendicular to this.
If there is potential however, a principal axis that respects that is desirable. 
For the special multi-HO potential problem, this has D as principal axis, being a permutation of the 
preceding. 
They have the feature that their projected quantum number is purely a relative angular momentum, so I 
write j in place of r.   
For the general multi-HO, the principal axis is in general unaligned with any kinematical features.  
See \cite{+tri} for how to proceed in that case.

%========================================================================================================
\subsection{Very special HO case $B = C = 0$ in spherical polar coordinates}
%========================================================================================================

I approach the QM via the correspondence with the mathematics for the Kepler--Coulomb problem that I 
pointed out in \cite{08I} and Sec 3.3. 
It corresponds to positive spatial curvature dust cosmology.  
This gives us the same wave equations as for the atomic problem and thus the usual separation and 
solvability in spherical and parabolic coordinates.  
Contrasting with \cite{ScaleQM}, there, the Coulomb problem is approximate and subject to 
small angle approximation breakdown, while in the present paper it is an exact solution.
On the other hand the present paper's analogy breaks down at the level of having a different inner 
product from the hydrogen problem.

Hydrogen has principal, angular momentum and magnetic quantum numbers ($\mn$, $\ml$, $\mm$ 
respectively), while the current system has analogues of these.  
Here, relative rational momentum quantum numbers r and R playing the roles of magnetic and total angular 
momentum quantum numbers.  
I use $\mN$ for the new principal quantum number that is associated with total moment of inertia of the 
system, which takes values such that ($K_1 = K_2 := K$ for the very special case) $K = E^2/\hbar^2\mN^2 
= 4\hbar^2/\mI_0^2\mN^2$ for $\mN \in \mathbb{N}$, Here, $4\hbar^2/E = I_0 = \mbox{Size}_0$ (the 
meaning of which is explained in Sec \ref{SSec: CharSc}).   
Thus one requires $E/\sqrt{K} = \mN\hbar \mbox{ } , \mbox{ } \mN \in \mathbb{N}$ as a consistency 
condition on the universe-model's energy and contents.

The corresponding wavefunctions are of the form 
\beq
\check{\Psi}_{\sN\sR\,\sj}(\mI, \Theta, \Phi) \propto \mL_{\sN - \sR - 1}^{2\sR + 1}({2\mI}/{\mN\mI_0}) 
\mbox{exp}(-{\mI}/{\mN \mI_0})({\mI}/{\mI_0})^{\sR}\mP_{\sR}^{\sj}(\mbox{cos}\,\Theta)
\mbox{exp}(i\mj\Phi) \mbox{ } , 
\eeq
for $\mP_{\sR}^{\sj}$ the associated Legendre functions and $\mL_{\alpha}^{\beta}$ the associated 
Laguerre polynomials.  
Then if the spherical polars are interpreted as a cluster-following basis, in terms of straightforward 
relational variables, 
\beq
\check{\Psi}_{\sN\sR\,\sj}(I_1, I_2, \Phi) \propto \mL_{\sN - \sR - 1}^{2\sR + 1}
\left(
\frac{E(I_1 + I_2)}{2\mN\hbar^2}
\right) 
\mbox{exp}
\left(
-\frac{E(I_1 + I_2)}{4\mN\hbar^2}
\right)
(I_1 + I_2)^{\sR}
\mP_{\sR}^{\sj}
\left(
\frac{I_2 - I_1}{I_1 + I_2}
\right)
\mbox{exp}(i\mj\Phi) \mbox{ } .   
\eeq
In terms of size and shape quantities, this is  
\beq
\check{\Psi}_{\sN\sR\,\sj}(\mbox{Size}, \mbox{aniso}, \mbox{ellip}) \propto 
\mL_{\sN - \sR - 1}^{2\sR + 1}
\left(
\frac{2\mbox{Size}}{\mN\,\mbox{Size}_0}
\right) 
\mbox{exp}
\left(
-\frac{\mbox{Size}}{\mN\,\mbox{Size}_0}
\right)
\left(
\frac{\mbox{Size}}{\mbox{Size}_0}
\right)^{\sR}
\mP_{\sR}^{\sj}
(\mbox{ellip})
{\cal T}_{\sj}
\left(
\mbox{aniso}/\sqrt{1 - \mbox{ellip}^2}
\right) \mbox{ } ,  
\eeq
where we have now taken sine and cosine combinations and ${\cal T}_{\sj}(\xi)$ means $T_{\sj}(\xi)$ for 
cosine solutions and $\sqrt{1 - T_{\sj}(\xi)^2}$ for sine solutions, for $T_{\sj}$ the Tchebychev 
polynomials. 
Also note that the Legendre variable here has the interpretation as the ellip shape variable.

I comment on the differences that the unusual inner product used here makes to the probability 
densities of various wavefunctions in Fig \ref{08III-Fig-4}.
Thus this paper's analogy is less extensive than \cite{08I, 08II}'s linear rigid rotor analogy.    
In this way the inclusion of scale complicates matters. 
Once nontrivial banal conformal transformations are needed, these have more implications for a 3-$d$ 
configuration space like the present paper's rather than for the 2-$d$ configuration spaces like in 
\cite{08II, +tri}, for which a number of cancellations occur.  
I sketch the first few shape states \cite{08II, +tri} in Fig 5.

One could proceed to investigate (as in \cite{08II}) the special case in spherical coordinates  
asymptotically and as a perturbation around the very special case.  
However, I have found an alternative exact method (c.f. the next 2 SSecs).

%FFFFFFFFFFFFFFFFFFFFFFFFFFFFFFFFFFFFFFFFFFFFFFFFFFFFFFFFFFFFFFFFFFFFFFFFFFFFFFFFFFFFFFFFFFFFFFFFFFFFFFFF 
{\begin{figure}[ht]
\centering
\includegraphics[width=0.8\textwidth]{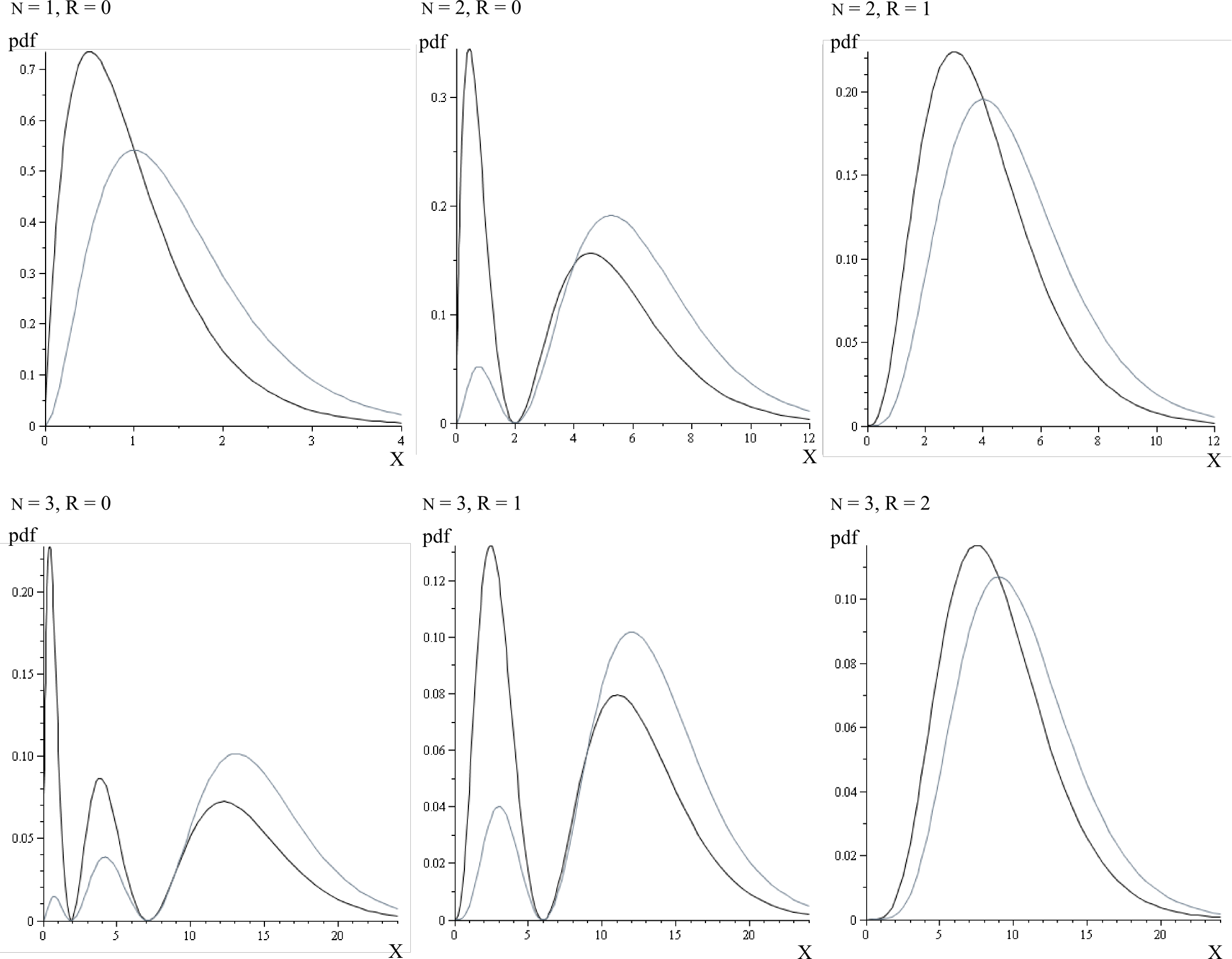}
\caption[Text der im Bilderverzeichnis auftaucht]{\footnotesize{How the probability density functions 
(pdf's) of this paper (black) compare to the hydrogenic ones (grey).  
These are plotted using Maple \cite{Maple} in terms of a dimensionless variable $X$, which is $I/I_0$ 
for the present paper's case and $r/a_0$ for the hydrogenic case.
Note that hydrogen's inner product is more suppressive closer to the origin than the present paper's. 
This furthermore means that rather more of the present paper's `$s^2$' pdf is 
inside the `$s^1$' one than in hydrogen, and likewise for the inner lobes of the `$s^3$' and `$p^3$' 
orbitals.}} \label{08III-Fig-4}\end{figure} } 
%FFFFFFFFFFFFFFFFFFFFFFFFFFFFFFFFFFFFFFFFFFFFFFFFFFFFFFFFFFFFFFFFFFFFFFFFFFFFFFFFFFFFFFFFFFFFFFFFFFFFFFFF

%FFFFFFFFFFFFFFFFFFFFFFFFFFFFFFFFFFFFFFFFFFFFFFFFFFFFFFFFFFFFFFFFFFFFFFFFFFFFFFFFFFFFFFFFFFFFFFFFFFFFFFF
{            \begin{figure}[ht]
\centering
\includegraphics[width=0.6\textwidth]{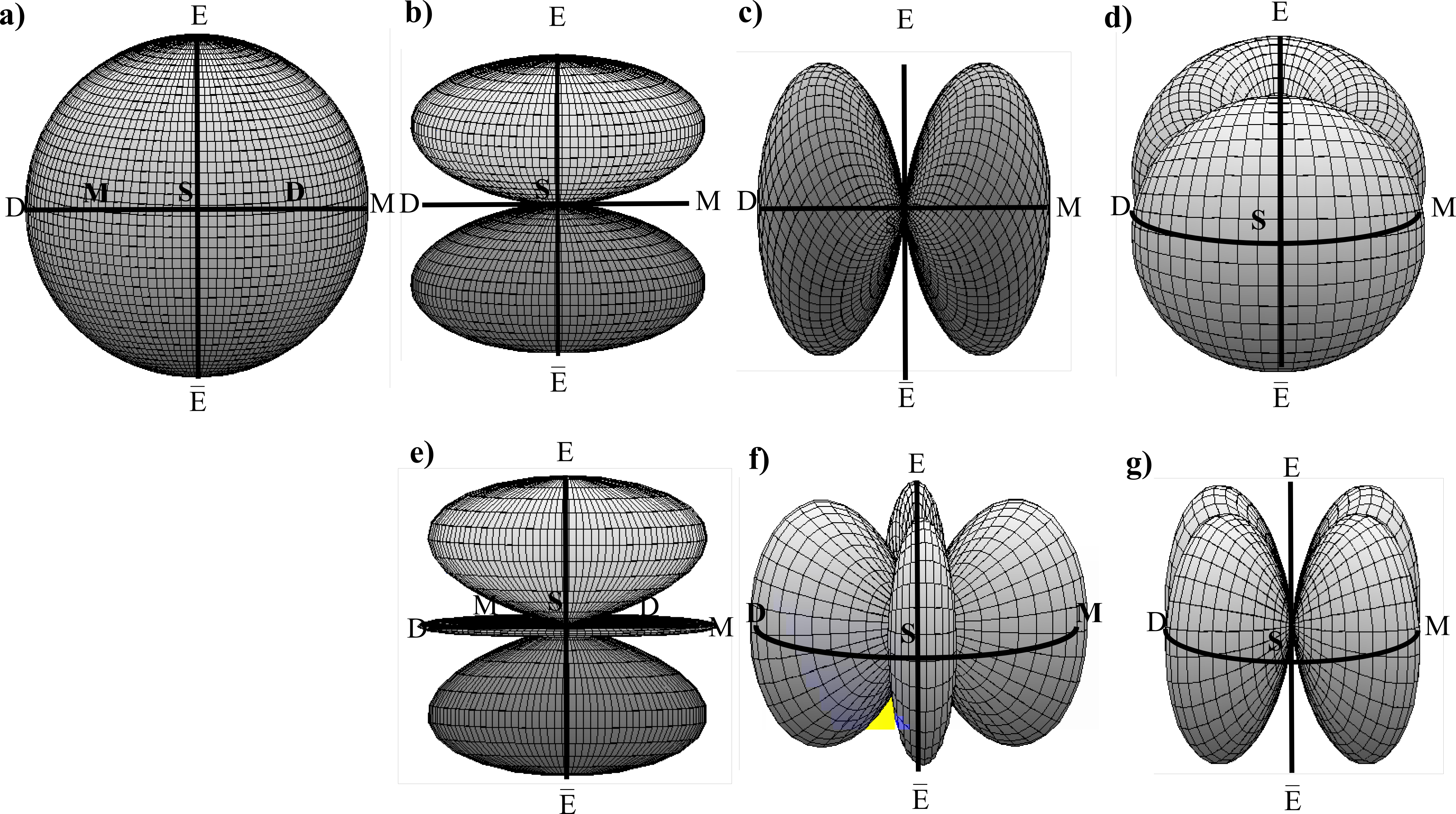}
\caption[Text der im Bilderverzeichnis auftaucht]{        \footnotesize{
\noindent An indication of which shapes are most probable for the first few states in the natural 
E-as-principal axis basis.  
The horizontal line is the collinear configurations, the vertical line is the regular configurations 
of the clustering that fixes the second axis. 
The bounding circle in the plane is the 
isosceles triangles corresponding to this clustering.  
a) is the ground state, that favours each place on the sphere equally. 

b), c), d) are first excited states.

e), f) and g) are three of the second excited states.   
The other two of these do not particularly favour anything that is geometrically significant.}      }
\label{Fig5}\end{figure}          }
%FFFFFFFFFFFFFFFFFFFFFFFFFFFFFFFFFFFFFFFFFFFFFFFFFFFFFFFFFFFFFFFFFFFFFFFFFFFFFFFFFFFFFFFFFFFFFFFFFFFFFFF

%========================================================================================================
\subsection{Free problem in parabolic-type coordinates}
%========================================================================================================

Solving (\ref{xiano}) in this case maps to the Bessel equation, giving an overall solution  
\beq
\check{\Psi}(\rho_1, \rho_2, \Phi) \propto 
\mJ_{\sj}(\rho_1/\rho_1(0))\mJ_{\sj}(\rho_2/\rho_2(0))\mbox{exp}(i\mj\Phi)
\eeq
for $\rho(0)^{e} = \hbar/\sqrt{2E_{e}}$.   
The interpretation of the probability density functions has the usual multiplicity of peaks of a free 
problem. 
These now correspond to a sequence of (cluster) separations that are probable alternating with a 
sequence that are improbable.
As is usual, have non-normalizability; in any case prefer models with potential terms 
(the latter is probably more primary, and non-normalizability is then a consequence of not doing so).

%===========================================================================================================================================
\subsection{Very special and special HO problems in parabolic-type coordinates}
%===========================================================================================================================================

The very special case matches the atomic problem in e.g. \cite{LLQM, Merzbacher, Hecht, Robinett}, under 
the correspondence $2I_{e} = \xi_{e}$, ${\cal E}$ to $-K/8$ (which is the right sign to get the 
bound states) and $e^2/4\pi\epsilon_0$ to $E$.  
Now, the quantum numbers are given by 
$
\mN_{\barp} = -(|\mj| + 1)/2 + \mN \beta_{\barp} 
$
for $\mN$ as before, $\mN_{\barp}$ parabolic quantum numbers and $\beta_{\barp}$ constants such that 
$\beta_1 + \beta_2 = 1$. 
Here 
\beq
\mN_1 + \mN_2 + |\mj| + 1 = \mN \mbox{ } . \label{sphepara}
\eeq
The time-independent Schr\"{o}dinger equation separates into simple harmonic motion in $\Phi$.  
While, for the other variables, one gets the same mathematics as for the standard atomic 
separated-out 1-d parabolic coordinate problem, 
\beq
\frac{1}{2}\frac{\d}{\d\mI_{e}}\left( \mI_{e} \frac{\d}{\d\mI_{e}}\Xi_{e} \right) - 
\frac{m^2}{8\mI_{e}}\Xi_{e} - \frac{K\mI_{e}}{8\hbar^2}\Xi_{e} = - 
\beta_{e}\frac{E}{2\hbar^2}\Xi_{e} \mbox{ } .   
\eeq
Thus the wavefunctions for the present problem up to normalization are (see e.g. \cite{LLQM, Hecht}) 
\beq
\check{\Psi}_{\sN_1\sN_2\,\sj}(I_1, I_2, \Phi) \propto 
\mL_{\sN_1}^{|\sj|}(2I_1/\mN I_0)
\mL_{\sN_2}^{|\sj|}(2I_2/\mN I_0)
\mbox{exp}
\left(
-(I_1 + I_2)/{\mN I_0}
\right)
(I_1I_2)^{|\sj|/2}\mbox{exp}(i\mj\Phi) \mbox{ } .  
\eeq

Next, the special case is almost as straightforward as the very special one.  
The separation continues to work out as before except that what is the same energy constant in each 
separated out parabolic problem now takes a different value for each, 
$-K^{\prime}_{e}/8$.\footnote{The 
%%%%%%%%%%%%%%%%%%%%%%%%%%%%%%%%%%%%%%%%%%%%%%%%%%%%%%%%%%%%%%%%%%%%%%%%%%%%%%%%%%%%%%%%%%%%%%%%%%%%%%%%
counterpart of this unusual extension was already remarked upon at the classical level in \cite{08I}.}  
%%%%%%%%%%%%%%%%%%%%%%%%%%%%%%%%%%%%%%%%%%%%%%%%%%%%%%%%%%%%%%%%%%%%%%%%%%%%%%%%%%%%%%%%%%%%%%%%%%%%%%%%
%
Now the quantum numbers come out to be as before, though each $\mN_{e}$ now has a distinct form of 
$\mN$: 
$\mN_{(e)} = 2\sqrt{2}\hbar/I\omega_{e}$ for $\omega_{e} = \sqrt{K_{e}}$,
so that there is not a simple relation like (\ref{sphepara}), but, rather, 
\beq
[\mN_1 + (|\mj| +  1)/2]/\mN_{(1)} + [\mN_2 + (|\mj| +  1)/2]/\mN_{(2)} = 1  \mbox{ } . 
\eeq
This corresponds to this less symmetric case not having a principal quantum number analogue. 
Its wavefunctions are  
\beq
\check{\Psi}_{\sN_1\sN_2\,\sj}(I_1,I_2, \Phi) \propto 
\mL_{\sN_1}^{|\sj|}(2I_1/\mN_{(1)}I_0)
\mL_{\sN_2}^{|\sj|}(2I_2/\mN_{(1)}I_0)
\mbox{exp}(- (I_1/\mN_{(1)} + I_2/\mN_{(2)})/I_0)
(I_1 I_2)^{|\sj|/2}\mbox{exp}(i\mj\Phi) \mbox{ } .  
\label{Lynx}
\eeq
Note that compared to the hydrogenic case, the difference in inner product causes the probability 
density functions for these to behave differently for m = 0/j = 0. 
(Our problem's then peak at the origin, while hydrogen's go to zero there).  

\noindent Also note that the significance of j for parabolic orbitals is as follows. 
j = 0 is the same probability for all relative angles. 

\noindent 
j = 1 has one state favouring the near-collinear configurations and another state favouring the 
near-right configurations. 
j = 2 has one state favouring both of these and one state favouring neither of them.

%========================================================================================================
\subsection{General HO problem in parabolic-type coordinates}
%========================================================================================================

Take (\ref{Lynx}) with normal coordinate $N$-labels on it and then apply the rotation in Sec 5 of 
\cite{08I}.  
Then, in a basis with principal axis aligned with the potential and E as second axis,  

\noindent
$$
\check{\Psi}_{\sN_1\sN_2\,\sr}(\mbox{Size}, \mbox{aniso}, \mbox{ellip}) =
\mL_{\sN_1}^{|\sr|}(\omega_1^{\sN}\mbox{Size}(f - B\, \mbox{ellip} - C\, \mbox{aniso})/g) 
\mL_{\sN_2}^{|\sr|}(\omega_2^{\sN}\mbox{Size}(f + B\, \mbox{ellip} + C\, \mbox{aniso})/g) 
\mbox{Size}^{|\sr|} \mbox{ } \times
$$
\beq 
(f^2-(B\,\mbox{ellip}+C\,\mbox{aniso})^2)^{|\sr|/2}\mbox{exp}(-\mbox{Size}
(f(\omega_1^{\sN}+\omega_2^{\sN})+(\omega_2^{\sN}-\omega_1^{\sN})(B\,\mbox{ellip}+C\,\mbox{aniso}))/g) 
{\cal T}_{\sr}
\left(
\frac{C\,\mbox{ellip}-B\,\mbox{aniso}}{\sqrt{f^2-(B\,\mbox{ellip}+C\,\mbox{aniso})^2}}
\right)
\eeq

\noindent
for $\omega_i^{\sN} = \sqrt{K_i^{\sN}}$ (normal mode frequencies), $f = \sqrt{B^2 + C^2}$ and 
$g = 2\hbar\sqrt{B^2 + C^2}$.

By comparison, recasting (\ref{Lynx}) in this scale--shape notation for the special case, 
$$
\Psi_{\sN_1\sN_2\,\sj}(\mbox{Size}, \mbox{aniso}, \mbox{ellip}) =
\mL_{\sN_1}^{|\sj|}(\omega_1\mbox{Size}(1 - \mbox{ellip})/2\hbar) 
\mL_{\sN_2}^{|\sj|}(\omega_2\mbox{Size}(1 + \mbox{ellip})/2\hbar)
\mbox{exp}(    -\mbox{Size}((\omega_1 + \omega_2) + (\omega_2 - \omega_1)\mbox{ellip})/4\hbar) 
$$
\beq
\times \mbox{ } \mbox{Size}^{|\sj|}(1 - \mbox{ellip}^2)^{|\sj|/2}
{\cal T}_{\sj}\left({\mbox{aniso}}/{\sqrt{1 - \mbox{ellip}^2}}\right) \mbox{ } . 
\eeq

%========================================================================================================
\subsection{Cases with further cosmology-inspired potentials}
%========================================================================================================

Some potentials that are tractable by methods parallel to those in \cite{ScaleQM} are 1) the 
upside-down HO's that map to ionized atoms, and correspond to negative spatial curvature in Cosmology. 
2) Extra $1/I^2$-type potentials, which allow for wrong-sign radiation to approximately become 
right-sign radiation. 
3) $I^2$-type potentials, corresponding to either sign of cosmological constant terms (Sec 6.3 covers 
this as a perturbation about this paper's main hydrogen-analogue model).

%========================================================================================================
%========================================================================================================
\section{Interpretation}
%========================================================================================================
%========================================================================================================

%=========================================================================================================
\subsection{Bohr moment of inertia}\label{SSec: CharSc}
%========================================================================================================

The mathematical analogy implies the following. 
\beq
 \mbox{ } \mbox{ (Bohr radius) } a_0 = 4\pi\epsilon_0\hbar^2/m_e e^2 
\mbox{ } \longleftrightarrow \mbox{ }  
\mI_0 = {4\hbar^2}/{E} \mbox{ } \mbox{ (a new `Bohr total moment of inertia' scale) } 
\eeq
which has the same kind of interpretation as the typical minimum quantity or effective size -- the 
overall ground state has a moment of inertia distribution with `characteristic width' $\mI_0$.  
N.B. The analogy with hydrogen at the quantum level is part-false, since the inner products do not match 
up.

%=========================================================================================================
\subsection{Expectations and spreads}
%========================================================================================================

As well as characterization by `modes and nodes' as are evident from figures such as Fig 4, 
expectations and spreads of powers of $r$ are used in the study of atoms. 
(See e.g. \cite{Messiah} for 
elementary use in the study of hydrogen, or \cite{FF} for use in approximate studies of larger atoms).   
These provide further information about the probability distribution function from that in the 
also-studied `modal' quantities (peaks and valleys) that are read off from plots or by the calculus.   
E.g. for hydrogen, from the angular factors of the integrals trivially cancelling and orthogonality and 
recurrence relation properties of Laguerre polynomials in \cite{AS} for the radial factors, one obtains 
the following.
E.g. in the $\ml = 0$ case,  
\beq
\langle\mn\,\ml\,\mm\,|\,r\,|\,\mn\,\ml\,\mm\rangle = (3\mn^2 - \ml(\ml + 1))a_0/2 
\mbox{ } \mbox{ and } \mbox{ } 
\Delta_{\sn\,\sll\,\sm}r = 
\sqrt{(\mn^2(\mn^2 + 2) - (\ml(\ml + 1))^2)}a_0/2 \mbox{ } ,
\eeq
where $a_0$ is the Bohr radius of the atom.
One can then infer from this that a minimal typical size is $3a_0/2$ and that 
the radius and its spread both become large for large quantum numbers.  
C.f. how the modal estimate of minimal typical size is $a_0$ itself. 
The slight disagreement between these is some indication of the limited accuracy to which either 
estimate should be trusted. 
Also, the above can be identified as expectations of scale operators, and thereby one can next ask 
whether they have pure shape counterparts in the standard atomic context.

Now, in spherical coordinates for the very special case, by the shape-scale split, and the 
orthogonality relation and a recurrence relation for the generalized Laguerre polynomials, 
\beq
\langle \mN\,\mR\,\mj\,|\,I\,|\,\mN\,\mR\,\mj \rangle = I_0\mN^2 \mbox{ } , \Delta_{\sN\,\sR\,\sj} = 
\sqrt{ \langle \mN\,\mR\,\mj\,|\,I^2\,|\,\mN\,\mR\,\mj\rangle - \langle \mN\,\mR\,\mj\, |\, I \, 
|\,\mN\,\mR\,\mj \rangle^2 } = I_0\mN\sqrt{\mN^2 + \mR(\mR - 1)}/\sqrt{2} \mbox{ } . 
\eeq
Thus, in particular for the ground state,  the expectation is the characteristic Bohr moment of inertia 
$I_0$ and the spread is $I_0/\sqrt{2}$.  
Differences between triangleland and hydrogen's expectations and spreads in r are due to the difference 
in inner product, noting that the present paper's case is easier to compute. 
(It requires just one use of the recurrence relation. 
Also, unlike for the atom, the expectation for triangleland turns out not to depend on the total 
rational quantum number.)

On the other hand, in parabolic coordinates, for the special case, 
\beq
\langle \mN_1\,\mN_2\,\mj\, |\, I \,|\, \mN_1\,\mN_2\,\mj \rangle = 
I_0\big(\mN_{(1)}\mN_1 + \mN_{(2)}\mN_2 + (\mN_{(1)} + \mN_{(2)})(|\mj| + 1)/2\big) 
\mbox{ } , \mbox{ } 
\Delta_{\sN_1\,\sN_2\,\sj} = 
I_0\mN\sqrt{  \mN_1^2 + \mN_2^2 + (\mN - |\mj|)( |\mj| + 1 ) }/\sqrt{2} 
\mbox{ } . 
\eeq
Note that for very special case ground state, these reduce to the above results, as they should.

%=========================================================================================================
\subsection{Perturbative treatment of additional cosmologically-inspired potential terms}
%========================================================================================================

E.g. $|{\bf r}_{IJ}|^6$ terms are cosmology-motivated, by corresponding to cosmological constant terms. 
These can be treated as small perturbations about the preceding Sec's problem.  
The perturbation theory expands in a series of powers of $\sigma$, so then there are also $O(\sigma^2)$ 
terms [and the previously mentioned corrections become $O(S^{u}, \sigma)$].    
They have a lead scale part $\sigma I^2$ and then $O(S^{u})$ terms. 
Then, using the same $\langle \mN\,\mR\,\mj\,|\, I^2 |\, \mN\,\mR\,\mj  \rangle$ integral as in the 
preceding SSec one obtains the first-order perturbation correction to the analogue of ${\cal E}$ to be 
$K = 4\hbar^2/\mN^2I_0^2 + SI_0^2\mN^2(3\mN^2 + \mR(\mR - 1))/8 + O(S^{u}, \sigma) + O(\sigma^2)$, 
whereupon an approximate inversion to look at the effect on $E$ gives that 
$\fE = \mN\hbar\omega - S(\hbar\omega)^3\mN(3\mN^2 + \mR(\mR - 1)) + O(S^{u}, \sigma) + O(\sigma^2)$. 
Thus $\sigma > 0$ lowers $E$, while $\sigma \leq 0$ raises $E$.

%========================================================================================================
%========================================================================================================
\section{Conclusion}
%========================================================================================================
%========================================================================================================

%========================================================================================================
\subsection{Outline of this paper's results so far}
%========================================================================================================

I studied the scaled triangle using 1) mass-weighted Jacobi coordinates. 
2) Dragt-type coordinates that play the role of Cartesian coordinates, furthermore interpreted as 
shape quantities (anisoscelesness, ellipticity and four times the area).  
3) Spherical polar coordinates, that are aligned with the shape--scale split. 
4) Parabolic coordinates that are aligned with the split into subsystems.  
5) Tessellation of the shape space sphere and relational space $\mathbb{R}^3$.  
It was found that the moment of inertia $I$ rather than the `hyperradius" $\sqrt{I}$ plays the natural 
role of radius.  
This paper's study has been geared towards specific solutions at the quantum level, with HO-like and 
cosmologically-inspired potentials.  
In this I was furthermore aided by mapping the wavefunction to that of the atomic problem, albeit the  
inner product is then different.
I interpreted these wavefunctions in terms of a Bohr moment of inertia scale, `modes and nodes' against 
the backcloth of the tessellation, expectations and spreads, and with further cosmologically-inspired 
potential terms treated perturbatively.

%===========================================================================================================================================
\subsection{Reduced versus Dirac quantization}
%===========================================================================================================================================

The Dirac-type approach gives somewhat different results \cite{Forth} from the present paper's 
reduced approach.  
One can see this as coming from reduction and the DeWitt-type operator-ordering not commuting as 
procedures.
This feature is absent from scaled N-stop metroland (as is obvious from these having no constraints), 
and also from pure-shape N-stop metroland.  
Thus one usefulness of triangleland is in being the simplest RPM for which reduced and Dirac approaches 
give different answers.  
I would argue that the reduced case should be trusted more, since adding physically irrelevant variables 
should not have the power of altering the true quantum theory of a physical system.
Thus here the reduced case provides `inside knowledge' of a good physical choice of operator-ordering.
This would cast doubts on Dirac-type approach results e.g. of \cite{06I} in the RPM context. 
N.B. also this is a `midi'- rather than `mini'-superspace issue, which may account for why it has not, to my best 
knowledge, been remarked upon before.    
I will comment to the extent to which this is a new QM phenomenon, with reference to previous 
literature, in \cite{Forth}.

%========================================================================================================
\subsection{Hidden time strategy}
%========================================================================================================

Scaled RPM possesses a hidden time of dilational type: the Euler time $t^{\sE\su\sll\se\sr} = \sum_i
{\bf R}^i \cdot {\bf P}_i$ \cite{06II, SemiclI}.   
This is an analogue of York time in GR.   
In fact, this analogy is not exact, highlighting that there is a multiplicity of scale variables whose 
canonical conjugates can then serve as dilational hidden times \cite{Forth}.  
For RPM's and minisuperspace, the analogue of the Lichnerowicz--York equation \cite{York73} required 
to isolate the true Hamiltonian $H_{\st\sr\su\se}$ is algebraic.  
Its tractability varies with choice of scale variable. 
This study reveals that both the constant mean curvature lapse-fixing equation (of theoretical 
numerical relativity \cite{BS03} as well as of the hidden York time approach \cite{K92, I93}) and the 
Lagrange--Jacobi equation for particle mechanics (familiar from celestial mechanics). 
Moreover, this study reveals that the generalizations of these two important equations from 
widely different branches of physics are in fact closely inter-related \cite{Forth}.

%========================================================================================================
\subsection{Emergent semiclassical time strategy}
%========================================================================================================

The heavy h degree of freedom is $I$, the mass-weighted scale.
The present paper adds to this by including consideration of in which regions various semiclassical 
approximations actually apply, including what is here the scale-dominates-shape approximation 
(\ref{SSA1}, \ref{SSA2}).    
The approximate emergent time $t^{\se\sm}$ that the h-system provides is computed for RPM's in
\cite{Cones}. 
For the five classical solutions in Sec 3.4, respectively, one gets
$t^{\se\sm} = \mbox{arcsin}(\sqrt{-2S} I)/\sqrt{-2S}$,
$t^{\se\sm} = \mbox{arccosh}(\sqrt{2S}I)/\sqrt{2S}$,
$t^{\se\sm} = \sqrt{2/9E}  I^{3/2}$, 
$t^{\se\sm} = I^2/2(2R - {\cal T}\!\mo\mt)$ and
$t^{\se\sm} =\sqrt{I^2 - {\cal T}\!\mo\mt}$.  
N.B. all these expressions are globally monotonic bar in the first `Milne in AdS' case, for which one 
can only have monotonicity for an epoch. 
2) these expressions (and further examples in \cite{Cones}) are invertible to
$I = I(t^{\se\sm})$.   
Thus that one can then replace $I$-dependence in the subsequent $l$-equations by 
$t^{\se\sm}$-dependence: eq (\ref{TDSE2})'s $\hat{H}_{\sll}$ can be written as $\hat{H}_{\sll}(I, S^{u}) 
= \hat{H}_{\sll}(t^{\se\sm}, S^{u}) = \hbar^2D_{S}^2/I(t^{\se\sm})^2\hbar^2 + J(I(t^{\se\sm}), S^{u})$.

This simplifies further if one uses instead the `rectified time' 
$
t^{\sr\se\sc} = t^{\sr\se\sc}(0) + \int\d t^{\se\sm}/I(t^{\se\sm})^2 
$. 
For the above five examples, this is  
$t^{\sr\se\sc} = \mbox{const} - \mbox{cot}(  \sqrt{-2S}t^{\se\sm}  )/( \sqrt{-2S})^{3/2}$,    
$t^{\sr\se\sc} = \mbox{const} + \mbox{tanh}(\sqrt{2S}t^{\se\sm})/(2S)^{3/2}$,  
$t^{\sr\se\sc} = \mbox{const} - ({2}/{E})^{2/3}/{{3t^{\se\sm}}^{1/3}}$, 
$t^{\sr\se\sc} =  \mbox{const} + \mbox{ln}\,t^{\te\tm}/{2\sqrt{2R - {\cal T}\!\mo\mt}}$ and  
\noindent $t^{\sr\se\sc} =   \mbox{arctan}({t^{\se\sm}}/({\cal T}\!\mo\mt - 2R))/
({\cal T}\!\mo\mt - 2R) + \mbox{const}  $.  
This is also invertible for the current paper's specific examples and with no loss in monotonicity. 
By this simplification, the time-dependent Schr\"{o}dinger equation is 
\beq
i\hbar\pa |\chi\rangle/\pa t^{\tr\te\tc} = -(\hbar^2/2)D_S^2|\chi\rangle + 
\widetilde{J}(t^{\sr\se\sc}, S^u)|\chi\rangle \mbox{ } \mbox{ for } \mbox{ }  
\widetilde{J} = I^2(t^{\se\sm}(t^{\sr\se\sc}))^2 J \mbox{ } ,  
\eeq
which can be viewed as a $t^{\sr\se\sc}$-dependent perturbation of what is, in the $N$-stop metroland 
case, a well-known $t^{\sr\se\sc}$-dependent Schr\"{o}dinger equation (the usual one on the 
circle/sphere/hypersphere).
It is an eventual aim of this program to consider the possibility of a less approximate semiclassical 
scheme involving more tightly coupled QM perturbation of the Hamilton--Jacobi equation coupled to a 
more accurate QM wave equation. 
Perhaps this should be seen \cite{SemiclI, Forth} as (an extension of) the well-known Hartree--Fock 
approach to Atomic and Molecular Physics.

Throughout the above, useful checks of the semiclassical approach's assumptions and approximations 
follow from RPM's having some \cite{08II, AF, +tri, MGM, ScaleQM, Forth} examples that are 
ulteriorly exactly soluble.  
(I.e., exactly soluble by means outside those that are usually available for specific toy models of the 
semiclassical approach, which are, moreover, seldom available in minisuperspace.)
The present paper's models and corresponding N-stop metroland ones \cite{ScaleQM} allow for a more 
detailed consideration of the semiclassical approach's assumptions and approximations (c.f.  
\cite{Forth}).

%========================================================================================================
\subsection{Timeless strategies}
%========================================================================================================

\noindent As an example of use of the na\"{\i}ve Schr\"{o}dinger inerpretation,
$
\mbox{P(size of the universe is less than some $I$)} \propto
$ 

\noindent$
\int_{I^{\prime} \leq I}
|\Psi(I^{\prime},\Theta, \Phi)|^2 I^{\prime}\mbox{sin}\,\Theta\, \d I^{\prime} \,\d\Theta\,\d\Psi \propto 
\int_{I^{\prime} = 0}^{I}{F}^2(I^{\prime}) I^{\prime} \d I^{\prime}
$ 
which, for the special multi-HO ground state, gives proportionality to $\mbox{exp}(-2I/I_0)(1 - I/I_0) 
- 1$.  
I also refer to \cite{+tri} for shape-part na\"{\i}ve Schr\"{o}dinger interpretation results, those 
arising again by the shape-scale split and triviality of the scale parts of the working.  
Investigating the conditional probabilities interpretation with RPM's is also possible.

Furthermore RPM's are amenable examples for the study of records theory.  
Their positive-definite kinetic metrics furbish suitable notions of distance \cite{Forth} on 
configuration space, e.g. the spherical metric or composite objects containing it such as action 
(\ref{accio}).
There are then mathematically well-defined notions of localization in space, such as almost-collinear/ 
almost-equilateral \cite{Kendall, +tri} (parametrized in each case by some small angle).  
It is then of interest what is the information/entropy in a triangle.  
This has small particle number issues and requires a somewhat unusual ensemble as it is a closed 
universe (fixed $E$) but not necessarily fixed particle number (due to collisions/coalescences).
This remains work in progress.

%========================================================================================================
\subsection{Histories theory strategies}
%========================================================================================================

For the first time, I announce that RPM's such as the present paper's scaled triangleland model 
can be cast as both Hartle-type \cite{Hartle} and Isham--Linden-type \cite{IL} histories theories 
\cite{Forth}.  
Hartle-type histories theories are well-known to involve coarse-graining operators and decoherence 
functionals. 
Isham--Linden-type histories theories are also known as the {\it histories 
projection operator} (HPO) approach; the {\it generally-covariant histories} or {\it histories 
brackets} approaches are similar. 
Here, there are two notions of time that are considered to be distinct. 
On the one hand, there is a kinematical notion of time that labels the histories as sequences of events. 
On the other hand, there is a dynamical notion of time that is generated by the Hamiltonian.

The main application I see for this RPM work on histories theory is the combination of semiclassical, histories theory and 
records theory ideas (see e.g. \cite{H03}).  
Quadrilateralland is possibly more interesting here (it decomposes into disjoint nontrivial 
subsystems). 
However, it makes sense to deal with triangleland first, as it is simpler for a number of reasons.
(Some of these concern $\mathbb{CP}^1 = \mathbb{S}^2$ versus $\mathbb{CP}^2$, but also others that 
become apparent in work in progress \cite{QShape}).

%========================================================================================================
\subsection{Other quantum cosmological applications}
%========================================================================================================

RPM's semiclassical approach scheme are useful toy models of midisuperspace Quantum Cosmology models 
that investigate the origin of structure formation in the universe. 
(E.g. the Halliwell--Hawking model toward Quantum Cosmology seeding galaxy formation and CMB 
inhomogeneities).
Thus RPM's are valuable conceptually and to test whether we should or should not be {\sl qualitatively} 
confident in the assumptions and approximations made in such schemes.
The spherical presentation of triangeland is of limited use due to not extending to higher 
$N$-a-gonlands.  
However, one use for it is in allowing the shape part to be studied in $\mathbb{S}^2$ 
terms which more closely parallel the Halliwell--Hawking \cite{HallHaw} analysis of GR inhomogeneities 
over $\mathbb{S}^3$.  
This application is more fully investigated in \cite{Forth}.

%\mbox{ }  

RPM's are also a useful toy model for notions of uniformity that are of widespread interest in Cosmology.
This applies to good approximation to the present distribution of galaxies and to the CMB. 
Furthermore, there is the issue of whether there was a considerably more uniform quantum-cosmological 
initial state \cite{Penrose}. 
There are also related issues of uniformizing process and how the small perturbations observed 
today were seeded.
However, uniformity is a pure-shape notion, and so it has already been covered in \cite{+tri} (and continues to apply 
as the shape part of the present paper's scaled model).  
The most uniform configuration here is the equilateral triangle.  
Almost-equilaterality can be investigated by the na\"{\i}ve Schr\"{o}dinger interpretation and is 
a notion of locality on configuration space.  

%\mbox{ }

RPM's are also a toy model for robustness issues, i.e. whether ignoring some degrees of freedom 
substantially changes the outcome.  
This is along the lines of Kucha\v{r} and Ryan \cite{KR89} questioning whether the Taub model subcase 
sits stably inside the Mixmaster model as regards making QM predictions. 
(This is itself a toy model of whether studying minisuperspace might be fatally flawed due to omitting 
all of the real universe's inhomogeneous modes).  
This was found to be unstable.  
RPM counterparts of this are rather more straightforward to investigate. 

\mbox{ }

%========================================================================================================
\noindent{\bf Acknowledgments}
%========================================================================================================
%
\noindent I thank: Professors Chris Isham, Karel Kucha\v{r} and Gary Gibbons, Dr Julian Barbour, 
Miss Anne Franzen, Mr Sean Gryb, Mr Henrique Gomes and Dr Jonathan Oppenheim for discussions. 
Professors Malcolm MacCallum, Don Page, Reza Tavakol, Belen Gavela and Marc Lachi\`{e}ze-Rey, 
and Dr's Jeremy Butterfield and Alexei Grinbaum for support in the furthering of my career.  
My wife, Alicia, Amelia, Coryan, Lynnette, Emma, Joshua, Emily, Sophie, Sophie and Sally for keeping my 
spirits up. 
Peterhouse, Fqxi grant RFP2-08-05, and Universidad Autonoma de Madrid for funding at various 
stages of this project.

%=====================================================BIBLIOGRAPHY==========================================================================


\begin{thebibliography}{99}
%===========================================================================================================================================

\footnotesize

%%%%%%%%%%%%%%%%%%%%%%%%%%%%%%%%%% A O R M APPLICABILITY OF RPM's %%%%%%%%%%%%%%%%%%%%%%%%%%%%%%%%%%%%%%%

\bibitem{BB82}               J.B. Barbour and B. Bertotti, Proc. Roy. Soc. Lond. {\bf A382} 295 (1982). 

\bibitem{ERPM}               J.B. Barbour, in {\it Quantum Concepts in Space and Time} 
                             ed. R. Penrose and C.J. Isham (Oxford University Press, Oxford 1986);    
%       
                             L.\'{A} Gergely, Class. Quantum Grav. {\bf 17} 1949 (2000), gr-qc/0003064; 
%
                             L.\'{A} Gergely and M. McKain, Class. Quantum Grav. {\bf 17} 1963 (2000), 
                             gr-qc/0003065. 

\bibitem{B94I}               J.B. Barbour, Class. Quantum Grav. {\bf 11} 2853 (1994).

\bibitem{EOT}                J.B. Barbour, {\it The End of Time} (Oxford University Press, Oxford 1999).

\bibitem{06I}                E. Anderson, Class. Quantum Grav. {\bf 23} 2469 (2006), gr-qc/0511068.

\bibitem{TriCl}              E. Anderson, Class. Quantum Grav. {\bf 24} 5317 (2007), gr-qc/0702083.  

\bibitem{08I}                E. Anderson, Class. Quantum Grav. {\bf 26} 135020 (2009), arXiv:0809.1168.   

\bibitem{Cones}              E. Anderson, arXiv:1001.1112 (This is a preprint that I do not intend to 
                             publish).

\bibitem{ScaleQM}            E. Anderson, arXiv:1003.4034.

\bibitem{B03}                J.B. Barbour, Class. Quantum Grav. \textbf{20} 1543 (2003), gr-qc/0211021. 

\bibitem{SRPM}               J.B. Barbour, in {\it Decoherence and Entropy in Complex Systems 
                             (Proceedings of the Conference DICE, Piombino 2002)} ed. H. -T. Elze, 
                             Springer Lecture Notes in Physics 2003), gr-qc/0309089.

\bibitem{06II}               E. Anderson, Class. Quantum Grav. {\bf 23} 2491 (2006), gr-qc/0511069.

\bibitem{FORD}               E. Anderson, Class. Quantum Grav. {\bf 25} 025003 (2008), arXiv:0706.3934.

\bibitem{08II}               E. Anderson, Class. Quantum Grav. {\bf 26} 135021 (2009) gr-qc/0809.3523.

\bibitem{+tri}               E. Anderson, arXiv:0909.2439. 

\bibitem{AF}                 E. Anderson and A. Franzen, Class. Quantum Grav. {\bf 27} 045009 (2010), 
                             arXiv:0909.2436.

%\bibitem{Tpaper}             E. Anderson, ``Notions of Uniformity in Metroland Quantum Cosmology", forthcoming.    

\bibitem{Rovellibook}        C. Rovelli, {\it Quantum Gravity} (Cambridge University Press, Cambridge 
                             2004).  

\bibitem{RWR}                J.B. Barbour, B.Z. Foster and N. \'{O} Murchadha, Class. Quantum Grav. 
                             {\bf 19} 3217 (2002), gr-qc/0012089;
%
                             E. Anderson and J.B. Barbour, Class. Quantum Grav. {\bf 19} 3249 (2002), 
                             gr-qc/0201092;
% 
                              E. Anderson, Phys. Rev. {\bf D68} 104001 (2003),  gr-qc/0302035; 
                             ``Geometrodynamics: Spacetime or Space?"  
                             (Ph.D. Thesis, University of London 2004), gr-qc/0409123;   
%
                             Stud. Hist. Phil. Mod. Phys. {\bf 38} 15 (2007), gr-qc/0511070;
%
                             in ``Classical and Quantum Gravity Research", ed. M.N. 
                             Christiansen and T.K. Rasmussen (Nova, New York 2008), arXiv:0711.0285.  

\bibitem{Lan}                E. Anderson, in {\it General Relativity Research Trends, Horizons in World 
                             Physics} {\bf 249} ed. A. Reimer (Nova, New York 2005), gr-qc/0405022.   

\bibitem{Dirac}              P.A.M. Dirac, {\it Lectures on Quantum Mechanics} (Yeshiva University, 
                             New York 1964). 
 
\bibitem{FEPI}               E. Anderson, Class. Quantum Grav. {\bf 25} 175011 (2008), arXiv:0711.0288.

\bibitem{AORM}               I. Newton, {\it Philosophiae Naturalis Principia Mathematica} (1686 and later editions).  
                             For an English translation, see e.g I.B. Cohen and A. Whitman (University of California Press, Berkeley, 1999). 
                             In particular, see the Scholium on absolute motion therein;   
%
                             see e.g. {\it The Leibniz--Clark correspondence} ed. H.G. Alexander 
                             (Manchester University Press, Manchester 1956);   
%
                             E. Mach, {\it Die Mechanik in ihrer Entwickelung,  Historisch-kritisch dargestellt} (J.A. Barth, Leipzig 1883).    
                             An English translation is {\it The Science of Mechanics:  A Critical and Historical Account of its Development} 
                             (Open Court, La  Salle, Ill. 1960);   
%
                             Bishop G. Berkeley {\it The Principles of Human  Knowledge} (1710);  
                             {\it Concerning Motion (De Motu)} (1721);  
%
                             J.B. Barbour, {\it Absolute or Relative Motion? Vol 1: 
                             The Discovery of Dynamics} (Cambridge University Press, Cambridge 1989);      
%
                             {\it Mach's principle: From Newton's Bucket to Quantum Gravity}  
                             ed. J.B. Barbour and H. Pfister (Birkh\"{a}user, Boston 1995).


\bibitem{+Phil}                J.B. Barbour, in {\it Quantum Concepts in Space and Time} 
                               ed. R. Penrose and C.J. Isham (Oxford University Press, Oxford, 1986);   
%
                              J. Earman, {\it World Enough and Space-Time: Absolute versus Relational 
                              Theories of Space and Time} (MIT Press, Cambridge MA, 1989);
%
                              R.A. Rynasiewicz, J. Phil. {\bf 93} 279 (1996); 
%
                              J.B. Barbour, in  {\it The Arguments of Time} ed. J. Butterfield 
                             (Oxford University Press, New York 1999);  
%
                              C. Hoefer, Brit. J. Phil. Sci. {\bf 49} 451 (1998); 
%
                              S. Saunders, in {\it Revisiting the Foundations of Relativistic Physics} 
                              ed. J. Renn (Kluwer, Dordrecht 2002);  
%
                              New York Times book review, 
                              http://www.nytimes.com/books/00/03/26/reviews/000326.26saundet.html;
%
                              L. Smolin, in {\it Time and the Instant} 
                              ed. R. Durie (Clinamen Press, Manchester 2000), gr-qc/0104097; 
%
                              J.N. Butterfield, Brit. J. Phil. Sci. {\bf 53} 289 (2002), gr-qc/0103055;    
%
                              O. Pooley, http://philsci-archive.pitt.edu/archive/00000221/index.html;
                              Proc. Phil. Time Soc. (2003-4);  
%
                              O. Pooley and H.R. Brown, Brit. J. Phil. Sci. {\bf 53} 183 (2002);  
%                            
                              L Sklar, Proc. Phil. Time Soc. {\bf 6} 64 (2003-4).
  
\bibitem{BS89etc}            J.B. Barbour and L. Smolin, unpublished, dating from 1989; 
%
                             L. Smolin, in {\it Conceptual Problems of Quantum Gravity} ed. A. Ashtekar 
                             and J. Stachel (Birkh\"{a}user, Boston 1991);  
% 
                             C. Rovelli, p. 292 in {\it Conceptual Problems of Quantum 
                             Gravity} ed. A. Ashtekar and J. Stachel (Birkh\"{a}user, Boston 1991);   
%
                             S.B. Gryb, arXiv:0804.2900; 
%
                             Class. Quantum Grav. {\bf 26} (2009) 085015, arXiv:0810.4152;  
                             arXiv:1003.1973;
%
                             J.B. Barbour and B.Z. Foster, arXiv:0808.1223.

\bibitem{Banal}              E. Anderson, Class. Quant. Grav. {\bf 27} 045002 (2010), arXiv:0905.3357. 

\bibitem{BSW}                R. Baierlein, D. Sharp, and J.A. Wheeler, Phys. Rev. {\bf 126}, 1864 (1962). 

\bibitem{ABFKO}              E. Anderson, J.B. Barbour, B.Z. Foster, B. Kelleher and N. \'{O} Murchadha, Class. Quantum Grav {\bf 22} 1795 (2005), gr-qc/0407104.              

\bibitem{Mini}               C.W. Misner, Phys. Rev {\bf 186} 1319 (1969);  
%
                             C.W. Misner, in {\it Relativity (Proceedings of the Relativity 
                             Conference in the Midwest, held at Cincinnati, Ohio June 2-6, 1969)} 
                             ed. M. Carmeli, S.I. Fickler and L. Witten (Plenum, New York 1970);
%
                             M. Ryan, {\it Hamiltonian Cosmology} (Lecture Notes in Physics 13) (Springer, Berlin, 1972);  
%
                             J.B. Hartle and S.W. Hawking, Phys. Rev. {\bf D28} 2960 (1983); 
%
                             M. Bojowald, Living Rev. Rel. {\bf 8} 11 (2005), gr-qc/0601085.  

\bibitem{Magic}              C.W. Misner, in {\it Magic Without Magic: John Archibald 
                             Wheeler} ed. J. Klauder (Freeman, San Fransisco 1972).

\bibitem{HH83}               J.B. Hartle and S.W. Hawking, Phys. Rev. {\bf D28} 2960 (1983).   

\bibitem{Wiltshire}          D.L. Wiltshire, in {\it Cosmology: the Physics of the Universe} 
                             ed. B. Robson, N. Visvanathan and W.S. Woolcock 
                             (World Scientific, Singapore 1996), gr-qc/0101003.   

\bibitem{K92}                K.V. Kucha\v{r}, in {\it Proceedings of the 4th Canadian Conference on 
                             General Relativity and Relativistic Astrophysics}   
                             ed. G. Kunstatter, D. Vincent and J. Williams (World Scientific, Singapore 1992). 

\bibitem{B94II}              J.B. Barbour, Class. Quantum Grav. {\bf 11} 2875 (1994).

\bibitem{Paris}              E. Anderson, AIP Conf. Proc. {\bf 861} 285 (2006), gr-qc/0509054.

\bibitem{SemiclI}            E. Anderson, Class. Quantum Grav. {\bf 24} 2935 (2007), gr-qc/0611007.

\bibitem{Records}            E. Anderson, Int. J. Mod. Phys. {\bf D18} 635 (2009), arXiv:0709.1892;   
                             in {\it Proceedings of the Second Conference on Time and 
                             Matter}, ed. M. O'Loughlin, S. Stani\v{c} and D. Veberi\v{c} 
                             (University of Nova Gorica Press, Nova Gorica, Slovenia 2008), arXiv:0711.3174.   

\bibitem{K81}                K.V. Kucha\v{r}, in {\it Quantum Gravity 2: a Second Oxford Symposium} 
                             ed. C.J. Isham, R. Penrose and D.W. Sciama (Clarendon, Oxford 1981).

\bibitem{K91}                K.V. Kucha\v{r}, in {\it Conceptual Problems of Quantum 
                             Gravity}, ed. A. Ashtekar and J. Stachel (Birkh\"{a}user, Boston 1991). 
      
\bibitem{I93}                C.J. Isham, in {\it Integrable Systems, Quantum Groups and Quantum Field Theories}  
                             ed. L.A. Ibort and M.A. Rodr\'{\i}guez (Kluwer, Dordrecht 1993), gr-qc/9210011.

\bibitem{K99}                K.V. Kucha\v{r}, in \it The Arguments of Time\normalfont, ed. J. Butterfield (Oxford University Press, Oxford 1999).

\bibitem{Kieferbook}         See e.g. C. Kiefer, {\it Quantum Gravity} (Clarendon, Oxford 2004).  

\bibitem{Smolin08}           L. Smolin, Problem of Time Course (2008), available in video form at http://pirsa.org/C08003.

\bibitem{Forth}              E. Anderson, forthcoming.

\bibitem{York72}              J.W. York, J. Math. Phys. {\bf 13} 125 (1972). 

\bibitem{York73}              J.W. York, Phys. Rev. Lett. {\bf 28} 1082 (1972); 
%
                              J. Math. Phys. {\bf 14} 456 (1973).

\bibitem{BS03}               T.W. Baumgarte and S.L. Shapiro, Phys. Rept. {\bf 376} 41 (2003), 
                             gr-qc/0211028.

\bibitem{York74}              J.W. York, Ann. Inst. Henri Poincar\'{e} {\bf 21} 319 (1974).                           

\bibitem{ABFO}               E. Anderson, J.B. Barbour, B.Z. Foster and N. \'{O} Murchadha, Class. Quantum Grav. {\bf 20} 157 (2003), gr-qc/0211022.  
 
\bibitem{HallHaw}            J.J. Halliwell and S.W. Hawking, Phys. Rev. {\bf D31}, 1777 (1985).
    
\bibitem{MGM}                E. Anderson, for Proceedings of Paris 2009 Marcel Grossman Meeting, 
                             arXiv:0908.1983.

%\bibitem{SemiclIII}          E. Anderson, ``On the Semiclassical Approach to Quantum Cosmology", 
%                             forthcoming.   

\bibitem{HP86}                S.W. Hawking and D.N. Page,  Nucl. Phys. {\bf B264} 185 (1986).   

\bibitem{UW89}               W. Unruh and R.M. Wald, Phys. Rev. {\bf D40} 2598 (1989). 

\bibitem{PW83}               D. Page and W. Wootters, Phys. Rev. {\bf D27} 2885 (1983).  

\bibitem{GMH}                 M. Gell--Mann and J.B. Hartle Phys. Rev {\bf D47} 3345 (1993). 

\bibitem{H99}                 J.J. Halliwell, Phys. Rev. {\bf D60} 105031 (1999), quant-ph/9902008.  

%\bibitem{NOD}                 E. Anderson, ``What is the Distance Between Two Shapes?",  forthcoming.  

\bibitem{Hartle}              J.B. Hartle, in {\it Gravitation and Quantizations} ed. B. Julia and 
                              J. Zinn-Justin (North Holland, Amsterdam 1995), gr-qc/9304063. 

\bibitem{Thiemann}           T. Thiemann, {\it Modern Canonical Quantum General Relativity} (Cambridge University Press, Cambridge 2007).

\bibitem{H03}                J.J. Halliwell, in {\it The Future of Theoretical Physics and Cosmology  
                             (Stephen Hawking 60th Birthday Festschrift Volume)} ed. G.W. Gibbons, 
                             E.P.S. Shellard and S.J. Rankin (Cambridge University Press, Cambridge 2003), gr-qc/0208018.  
 
\bibitem{Dragt}              A.J. Dragt, J. Math. Phys. {\bf 6} 533 (1965). 
 
%\bibitem{FileR}              E. Anderson, forthcoming.  
%
%\bibitem{NOI}                 E. Anderson, forthcoming.  

%%%%%%%%%%%%%%%%%%%%%%%%%%%%%%%%% B O D Y   R E F E R E N C E S %%%%%%%%%%%%%%%%%%%%%%%%%%%%%%%%%%%%%%%%%

\bibitem{Marchal}             See e.g. C. Marchal, {\it Celestial Mechanics} (Elsevier, Tokyo 1990).

\bibitem{Kendall}             D.G. Kendall, D. Barden, T.K. Carne and H. Le, {\it Shape and Shape 
                              Theory} (Wiley, Chichester 1999).  

\bibitem{Mont98}             R. Montgomery, Nonlin. {\bf 11} 363 (1998), math/9510005.

\bibitem{ArchRat}            R. Montgomery, Arch. Rat. Mech. Anal. {\bf 164} 311 (2002). 

\bibitem{Hsiang}             W.-Y. Hsiang and E. Straume,  arXiv:math-ph/0609084; math-ph/0608060; math-ph/0609076.  

\bibitem{Zick}               W. Zickendraht, Phys. Rev. {\bf 159} 1448 (1967); 
%
J. Math. Phys. {\bf 10} 30 (1969);                 
%
J. Math. Phys. {\bf 12} 1663 (1970). 

\bibitem{ACG86}              V. Aquilanti, S. Cavalli and G. Grossi, J. Chem. Phys. {\bf 85} 1362 (1986).  

\bibitem{LR95}               R.G. Littlejohn and M. Reinsch, , Phys. Rev. {\bf A52} 2035 (1995).

\bibitem{Montgomery2}        R. Montgomery, Ergod. Th. Dynam. Sys. {\bf 25} 921 (2005), math/0405014.

\bibitem{GibGri}             G.W. Gibbons and L.P. Grischuk, Nu. Phys. {\bf B313} 736 (1989).  

\bibitem{Amsterdamski}       P. Amsterdamski, Phys. Rev. {\bf D31} 3073 (1985).  

\bibitem{CosMech}            E.A. Milne, Quart. J. Math. {\bf 5} 64 (1934);   
%
                             W.H. McCrea and E.A. Milne,  Quart. J. Math. {\bf 5} 73 (1934);
%
                             E.A. Milne, {\it Relativity, Gravitation and World Structure} 
                             (Oxford University Press, Oxford 1935).      

\bibitem{BGS03}              R.A. Battye, G.W. Gibbons and P.M. Sutcliffe, Proc. R. Soc. Lond. 
                             {\bf A 459} 911 (2003), hep-th/0201101. 

\bibitem{HarMTWPee}          E.R. Harrison, Ann. Phys. {\bf 35} 437 (1965); 
%
                             pp. 733-4 of C.W. Misner, K. Thorne and J.A Wheeler, {\it Gravitation} 
                             (Freedman, San Francisco 1973); 
%
                             P.J.E. Peebles, {\it Principles of Physical Cosmology} (Princeton University 
                             Press, Princeton 1993).  

\bibitem{Rindler}            W. Rindler, {\it Relativity. Special, General and Cosmological.} 
                             (Oxford University Press, New York 2001). 

\bibitem{DarkRad}            See e.g. N. Tetradis, J. Phys.: Conf. Ser. {\bf 68} 012034 (2007).  

\bibitem{AT05}               E. Anderson and R.K. Tavakol, JCAP 0510 017 (2005), gr-qc/0509055.   

\bibitem{Isham84}            C.J. Isham, in {\it Relativity, Groups and Topology {II}} ed. B.S. DeWitt and R. Stora 
                             (North-Holland, Amsterdam 1984). 

\bibitem{DeWitt57}           B.S. DeWitt, Rev. Mod. Phys. {\bf 29} 377 (1957).  

\bibitem{Oporder}            T. Christodoulakis and J. Zanelli, Nuovo Cim. {\bf B93} 1 (1986); 
%
                             J.J. Halliwell, Phys. Rev. {\bf D38} 2468 (1988);
%
                             I. Moss, Ann. Inst. H. Poincar\'{e} {\bf 49} 341 (1988); 
%
                             D.N. Page, J. Math. Phys. {\bf 32} 3427 (1991);
%
                             M.P. Ryan and A.V. Turbiner, Phys. Lett. {\bf A333} 30 (2004), quant-ph/0406167.

\bibitem{Wald}               R.M. Wald, {\it General Relativity} (University of Chicago Press, Chicago 1984).  

\bibitem{IwaiUwano}           T. Iwai and Y. Uwano, J. Math. Phys. {\bf 27} 1523 (1986); 
%
                              A. Tachibana and T. Iwai, Phys. Rev. {\bf A33} 2262 (1986);  
%
                              T. Iwai and Y. Iwano, J. Phys. A. Math. Gen. {\bf 21} 4083 (1988).  

\bibitem{Maple}              This was plotted using Maple 12.  

\bibitem{LLQM}                L.D. Landau and E.M. Lifshitz, {\it Quantum Mechanics} (Pergamon, New York 1965).

\bibitem{Merzbacher}          E. Merzbacher, {\it Quantum Mechanics} (Wiley, New York 1998). 

\bibitem{Hecht}               K.T. Hecht, {\it Quantum Mechanics} (Springer-Verlag, New York 2000).  

\bibitem{Robinett}            R.W. Robinett, {\it Quantum Mechanics: Classical Results, Modern Systems, and Visualized Examples} 
                              (Oxford University Press, New York 1997). 

\bibitem{Messiah}           A. Messiah, {\it Quantum Mechanics} Vol 2 (North Holland, Amsterdam 1965).  

\bibitem{FF}                C. Frose Fischer, {\it The Hartree-Fock Method for Atoms} (Wiley, New York 1977).  

\bibitem{AS}                M. Abramowitz and I.A. Stegun, {\it Handbook of Mathematical Functions} (Dover, New York 1970).     

\bibitem{IL}                  C.J. Isham and N. Linden, J. Math. Phys. {\bf 36} 5392 (1995), 
                              gr-qc/9503063. 

%\bibitem{AHist}             E. Anderson, ``History of a Triangle", forthcoming.

\bibitem{Penrose}           R. Penrose, {\it The Road to Reality} (Vintage, London 2005).  

\bibitem{KR89}              K.V. Kucha\v{r} and M.P. Ryan, Phys. Rev. {\bf D40} 3982 (1989).

\bibitem{QShape}            E. Anderson, arXiv:1009.2161.    

\bibitem{APOT}              E. Anderson, arXiv:1009.2157.

\end{thebibliography}
\end{document}